\documentclass[11pt]{article}
\usepackage{amsmath,amsthm,amsrefs,amstext,amssymb,amsbsy}
\usepackage{mathtools}
\usepackage{graphicx}
\usepackage{caption, subcaption}
\usepackage{fullpage}
\usepackage{color}
\usepackage[colorlinks=true, citecolor=blue]{hyperref}
\usepackage[usenames,dvipsnames]{xcolor}
\usepackage{array}
\usepackage{tikz}
\usepackage{kpfonts}

\newtheorem{lem}{Lemma}
\newtheorem{cor}{Corollary}
\newtheorem{thm}{Theorem}

\newtheorem{conj}{Conjecture}
\theoremstyle{remark}
\newtheorem*{rmk*}{Remark}
\theoremstyle{definition}
\newtheorem{assns}{Assumptions}

\newcommand{\edge}[1][1em]{\!\mathrel{\rule[0.5ex]{#1}{.4pt}}\!}
\newcommand{\smin}{s_\mathrm{min}}

\newcommand{\tube}{\mathbb T}
\newcommand{\tubeLM}{\mathbb T_{L,M}}
\newcommand{\bfull}{b^\mathrm{F}}
\newcommand{\betafull}{\beta^\mathrm{F}}
\newcommand{\pham}{p^\mathrm{H}}
\newcommand{\kappaham}{\kappa^\mathrm{H}}
\newcommand{\Zham}{Z^\mathrm{H}}
\newcommand{\Fham}{\mathcal F^\mathrm{H}}

\title{Polygons in restricted geometries subjected to infinite forces}
\author{N R Beaton, J W Eng and C E Soteros
\\Department of Mathematics and Statistics\\
University of Saskatchewan,  Saskatoon, SK, Canada S7N 5E6}

\begin{document}
\maketitle
\begin{abstract}
We consider self-avoiding polygons in a restricted geometry, namely an infinite $L\times M$ tube in $\mathbb Z^3$. These polygons are subjected to a force $f$, parallel to the infinite axis of the tube. When $f>0$ the force stretches the polygons, while when $f<0$ the force is compressive. We obtain and prove the asymptotic form of the free energy in both limits $f\to\pm\infty$.
 We conjecture that the $f\to-\infty$ asymptote is the same as the {limiting free energy of ``Hamiltonian'' polygons, polygons} which visit every vertex in a $L\times M\times N$ box.
 We investigate such polygons, and in particular use a transfer-matrix methodology to {establish that the conjecture is true for  some small tube sizes}.
\end{abstract}

\begin{center} Dedicated to Anthony J. Guttmann on the occasion of his 70\textsuperscript{th} birthday.\end{center}

\section{Introduction}\label{sec:intro}

Since the advent of single molecule experiments using, for example, atomic force microscopy, there has been much interest in modelling polymers subject to a tensile force (see for example~\cite{Faragoetal2002, Krawczyketal2005,vanRensburgetal2008stretched, Atapouretal2009Stretched, Ioffeetal2010review, Beaton2015, Beatonetal2015}).  Models range from random walk in $\mathbb{R}^3$  to lattice models and they have been studied both numerically and using combinatorial or probabilistic analysis.
Recent advances on the theoretical side, include a proof for the self-avoiding walk (SAW) lattice model of linear polymers that there is a phase transition  between a free and a ballistic phase at a critical force, $f_\mathrm{c}$, corresponding to when the force, $f=f_\mathrm{c}=0$ \cite{Beaton2015}.  Most recently, for the square lattice,  conjectures based on Schramm-Loewner evolution have been used to predict the form of the partition function and associated critical exponents \cite{Beatonetal2015}.

From the beginning, one particular area of focus has been on the effect of topological constraints \cite{Faragoetal2002} and, for example,  how the  knotting probability in ring polymers depends on the force \cite{vanRensburgetal2008stretched}.  For a lattice model of this,  self-avoiding polygons on the simple cubic
lattice are the standard model.   For this case,  Janse van Rensburg et al  \cite{vanRensburgetal2008stretched} found that for sufficiently large fixed forces, all but exponentially few sufficiently large
polygons are knotted.  It is believed that this should hold for any force $f$, but this has yet to be proved.  By restricting the polygons to lie in a lattice tube however, Atapour et al \cite{Atapouretal2009Stretched} proved that for any fixed force (either stretching or compressing),  all but exponentially few sufficiently large polygons are knotted.   The proof was based on transfer-matrix theory and pattern theorem arguments.
In this paper, we explore the Atapour et al model further by investigating the asymptotes as the force goes to either plus or minus infinity.   We establish the existence of
the asymptotes and their form. 
Furthermore,  we determine a subset of polygons whose free energy becomes dominant
 in the limit as the force goes to negative infinity.
 One subset of these polygons are those which correspond to undirected Hamiltonian circuits (called \textit{Hamiltonian} polygons);  {using arguments adapted from \cite{Eng2014Thesis}
 we establish for this subset
 that the limiting free energy exists, and we review the result from \cite{Eng2014Thesis} that all but exponentially few sufficiently large Hamiltonian polygons are knotted.} From transfer-matrix calculations, {we also explore whether Hamiltonian polygons dominate as the force goes to negative infinity. 
We establish that they do dominate for small tube sizes, and conjecture that this holds for all tube sizes. If this conjecture holds then, for example,  for any force $f\in[-\infty, \infty)$, all but  exponentially few sufficiently large polygons will be knotted.}

In this paper we use exact enumeration and transfer-matrix methods to study self-avoiding polygons, building on the numerous contributions of A. J. Guttmann to this area. For example, in~\cite{Enting_1985, Guttmann_1988},   Guttmann and collaborators developed
 transfer matrix methods for efficient exact enumeration to, amongst other things,  obtain bounds on growth constants and study the critical exponents for
 polygons on the square lattice.
 In the recent paper \cite{Beatonetal2015}, related approaches are used to study compressed walks, bridges and polygons.  Here we follow in a similar vein  but explore compressed and stretched three-dimensional polygons embedded in an essentially one-dimensional lattice subset and we use transfer-matrix theory and exact enumeration/generation methods to obtain relationships between free energies and growth constants.
 
The paper is structured as follows.  First the details of the Atapour et al model are reviewed, highlighting known upper and lower bounds for the free energy as a function of the force $f$.  Next we establish the asymptotic forms for the free energy, first as $f\to\infty$ and next as $f\to -\infty$.  Finally we prove results about Hamiltonian polygons and use transfer matrix arguments  for small tube sizes to validate our conjecture that they dominate the free energy as the force goes to minus infinity.

\section{The model}\label{sec:themodel}

\begin{figure}
\centering
%\begin{tikzpicture}
%[scale=0.5, every node/.style={circle, fill, inner sep=2.5pt}]
%\begin{knot}[clip width=3]
%\strand[ultra thick] (0,0) node{} -- (0,3) node{} -- (1,4) node{} -- (2,5) node{} -- (5,5) node{} -- (8,5) node{} -- (11,5) node{} -- (14,5) node{} -- (17,5) node{} -- (16,4) node{}  -- (16,1) node{} -- (15,0) node{} -- (12,0) node{} -- (9,0) node{} -- (9,3) node{} -- (6,3) node{} -- (3,3) node{} -- (4,4) node{};
%\strand[ultra thick] (4,4) -- (4,1) node{} -- (5,2) node{} -- (8,2) node{} -- (11,2) node{} -- (14,2) node{} -- (17,2) node{} -- (20,2) node{} -- (20,5) node{} -- (19,4) node{} -- (18,3) node{} -- (15,3) node{} -- (12,3) node{} -- (13,4) node{};
%\strand[ultra thick] (13,4) -- (13,1) node{} -- (10,1) node{} -- (7,1) node{} -- (6,0) node{} -- (3,0) node{} -- (0,0);
%\flipcrossings{2,6}
%\end{knot}
%\end{tikzpicture}
\includegraphics{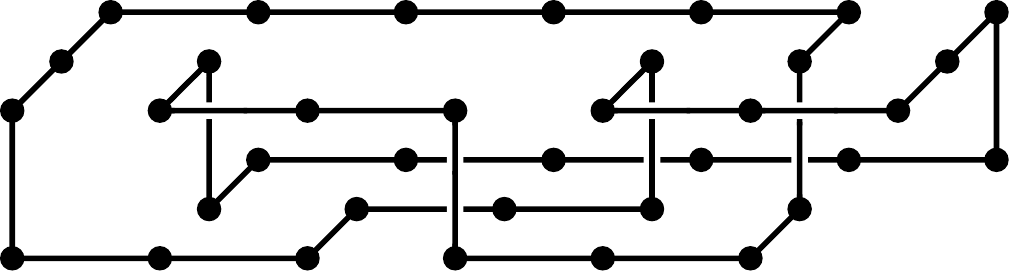}
\caption{A self-avoiding polygon in the $2\times1$ tube. This polygon has length 36 and span 6.}
\label{fig:polygon}
\end{figure}

{For non-negative integers $L,M$, let} \(\tubeLM\equiv\tube\subset \mathbb Z^3\) be the semi-infinite $L\times M$ tube on the cubic lattice defined by
\[\tube = \{(x,y,z)\in \mathbb Z^3:x\geq0, 0\leq y\leq L, 0 \leq z \leq M\}.\]
Define $\mathcal P_\tube$ to be the set of self-avoiding polygons in $\tube$ which occupy at least one vertex in the plane $x=0$, and let $\mathcal P_{\tube,n}$ be the subset of $\mathcal P_\tube$ comprising polygons with $n$ edges. Then let $p_{\tube,n}  = |\mathcal P_{\tube,n}|$. See Figure~\ref{fig:polygon} for a polygon in the $2\times1$ tube.

\begin{rmk*} Throughout the rest of this paper, the symbol $n$ will only be used to denote the number of edges in polygons. We will thus always assume that $n$ is even. This includes limits and, for example, $\lim_{n\to\infty}$ should be interpreted as a limit through even values of $n$ only.  {Furthermore, for $L=M=0$, $p_{\tube,n} =0$ for all $n$,  thus for the rest of the paper we assume at least one of $L$ or $M$ is strictly positive.}
\end{rmk*}

We define the \emph{span} $s(\pi)$ of a polygon $\pi\in\mathcal P_\tube$ to be the maximal $x$-coordinate reached by any of its vertices and we use $|\pi|$ to denote the number of edges in $\pi$. To model a force acting parallel to the $x$-axis, we  associate a fugacity (Boltzmann weight) $e^{fs(\pi)}$ with each polygon $\pi$. Let $p_{\tube,n}(s)$ be the number of polygons in $\mathcal P_{\tube,n}$ with span $s$. Then define the partition function
\begin{displaymath}
Z_{\tube,n}(f)  = \sum_{|\pi| = n} e^{fs(\pi)} = \sum_s p_{\tube,n}(s) e^{fs}.
\end{displaymath}

The weight $f$ represents a force in the following way: when $f\ll 0$, polygons with small span will dominate the partition function, so this corresponds to the ``compressed'' regime. On the other hand, when $f\gg 0$, polygons with large span will dominate the partition function, so this corresponds to the ``stretched'' regime.

We will use the notation $W=(L+1)(M+1)$ (the number of vertices in an integer plane $x=i\geq 0$ of the tube) for shorthand, and will assume without loss of generality that $L\geq M$.
Note that for any $n\geq 4$ the minimum span for any $n$-edge polygon, $\smin(n)$, is such that $p_n(\smin(n))>0$ and given any polygon $\pi\in\mathcal P_{\tube,n}$,  $s(\pi)\geq \smin(n)\geq \frac{n}{W}$.  The maximum span of an $n$-edge polygon is $\frac{n-2}{2}$ \cite{Atapouretal2009Stretched}.  We thus have the following bounds which correct~\cite{Atapouretal2009Stretched}*{eqn.~(6)}:
\begin{align}\label{eqnpartitionfunctionbounds}
\max\{e^{f(n-1)/2},p_{\tube,n}(\smin(n)) e^{f\smin(n)}\} &\leq Z_{\tube,n}(f) \nonumber \\
&=  \sum_s p_{\tube,n}(s) e^{fs} \nonumber\\
&\leq  \max\{e^{f\smin(n)},e^{f(n-1)/2}\}{ p_{\tube,n}}.
\end{align}

The \emph{free energy} of polygons in $\tube$ is defined as
\[\mathcal F_\tube(f) = \lim_{n\to\infty}\frac1n \log Z_{\tube,n}(f).\]
This is known \cite{Atapouretal2009Stretched} to exist for all $f$. It is a convex function of $f$, and is thus continuous and almost-everywhere differentiable.
It has been proved {\cite{Atapouretal2009Stretched} that:
\begin{equation}\label{eqn:Zn_asymp_simple_pole}
Z_{\tube,n}(f) ={ \alpha_{\tube}}(f) e^{\mathcal F_\tube(f)n}\left(1+O(n^{-1})\right),
\end{equation}
where $\alpha_{\tube}(f)$ depends only on $f$, $L$ and $M$. 
From this it also follows that, for example,
\begin{equation}\label{eqn:Zn_ratio}
\lim_{n\to\infty} \frac{Z_{\tube,n+2}(f)}{Z_{\tube,n}(f)} = e^{2\mathcal F_\tube(f)}.
\end{equation}

Note that $|\mathcal{P}_{\tube,n}|= p_{\tube,n}=Z_{\tube,n}(0)\leq Wp_n$, where $p_n$ is the number of $n$-edge self-avoiding polygons in $\mathbb Z^3$ counted up to translation.
It has been proved that {\cite{SotWhit88,Sot89},
\begin{equation}\label{eqnmudef}
\mathcal F_\tube(0)=\lim_{n\to\infty} n^{-1}\log p_{\tube,n}  < \lim_{n\to\infty} n^{-1}\log p_n =\lim_{n\to\infty} n^{-1}\log c_n \equiv \kappa \equiv \log \mu, 
\end{equation}
}
where $c_n$ is the number of $n$-step self-avoiding walks (SAWs) in $\mathbb Z^3$ starting at the origin and $\kappa$ is their connective constant.

The bounds in \eqref{eqnpartitionfunctionbounds} lead to the following bounds on the free energy:
\begin{align*}
\max\{f/2,(f/W)+\limsup_{n\to\infty} n^{-1}\log p_{\tube,n}(\smin(n)) \} &\leq \mathcal F_\tube(f)\\
&\leq  \max\{f/W,f/2\} + \mathcal F_\tube(0).
\end{align*}

For the lower bound, one set of polygons which have minimum span are the Hamiltonian polygons.  We define the number of Hamiltonian polygons, $\pham_{\tube,n}$, to be the number of  $n$-edge, for $n=W(s+1)$, span-$s$ polygons in $\mathcal{P}_{\tube,n}$ which occupy every vertex in an $L\times M \times s$ subtube of $\tube$.  In \cite{Eng2014Thesis}, the following limit is proved to exist and we have:
\begin{equation*}\label{eqn:ham_growthrate1}
\kappaham_\tube \equiv \lim_{s\to\infty}\frac{1}{(s+1)W} \log \pham_{\tube,(s+1)W}\leq \limsup_{n\to\infty} n^{-1}\log p_{\tube,n}(\smin(n)).
\end{equation*}
Thus another set of bounds for the free energy is given by:
\begin{align}
\max\{{f/2},(f/W)+ \kappaham_\tube\} &\leq \mathcal F_\tube(f) \leq  \max\{{f/W},{f/2}\} +\mathcal F_\tube(0).
\label{eqnfreeenergybounds}
\end{align}
For small tube sizes,  $\mathcal F_\tube(f)$, $f\in(-\infty, \infty)$, and $\kappaham_\tube$ have been obtained from numerical calculations of the eigenvalues of appropriate transfer matrices \cite{Eng2014Thesis}; the resulting free energy and bounds associated with \eqref{eqnfreeenergybounds} are shown in Figure \ref{fig:fe_plots} (more details about these calculations will be given in Section \ref{sec:hamiltonian}).  These graphs strongly suggest that the free energy is asymptotic to the lower bound as $f$ goes to $\pm\infty$. In the next section we explore this proposition, and prove that it is indeed the case for  $f\to\infty$.   We also establish the form  for the asymptote as $f\to -\infty$ and provide further evidence, for small tube sizes, that it corresponds to the lower bound in \eqref{eqnfreeenergybounds}.

\begin{figure}
\centering
\begin{subfigure}{0.48\textwidth}
\includegraphics[width=\textwidth]{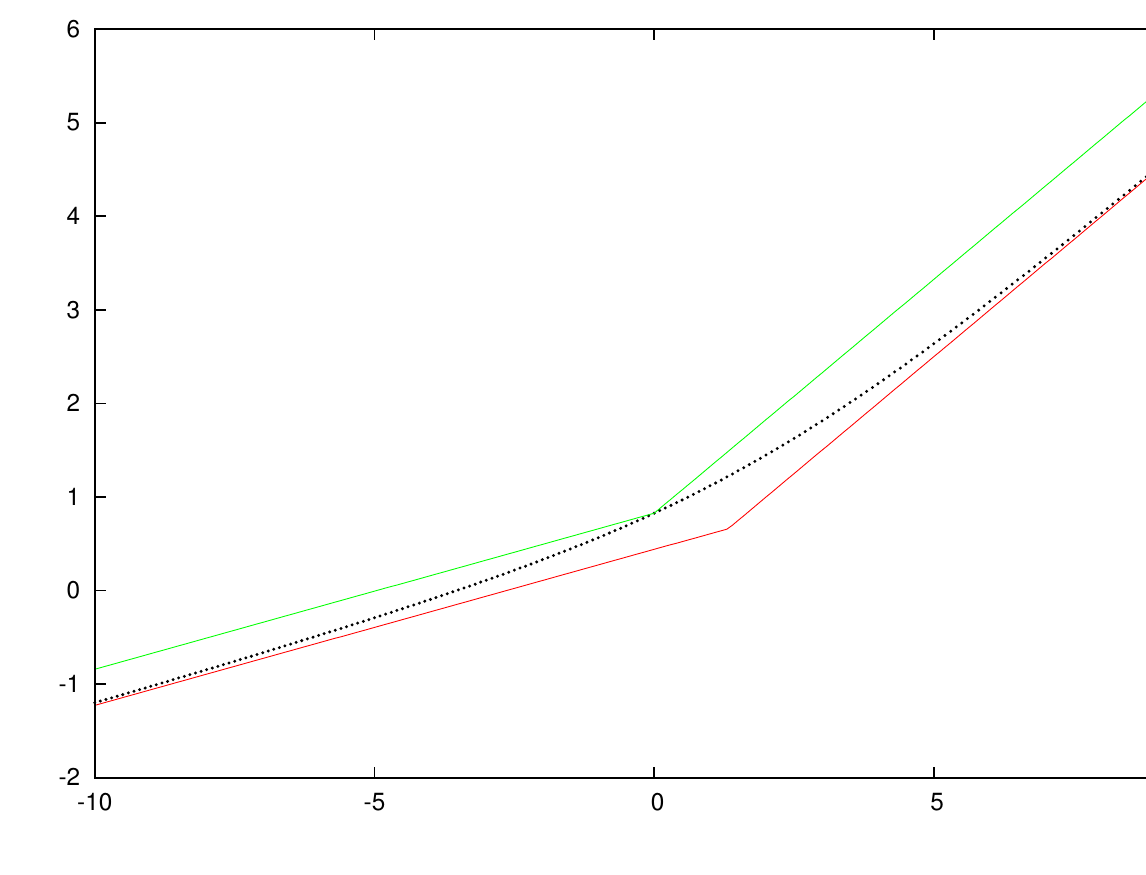}
\caption{Free energies in the $2\times1$ tube.}
\end{subfigure}
\hfill
\begin{subfigure}{0.48\textwidth}
\includegraphics[width=\textwidth]{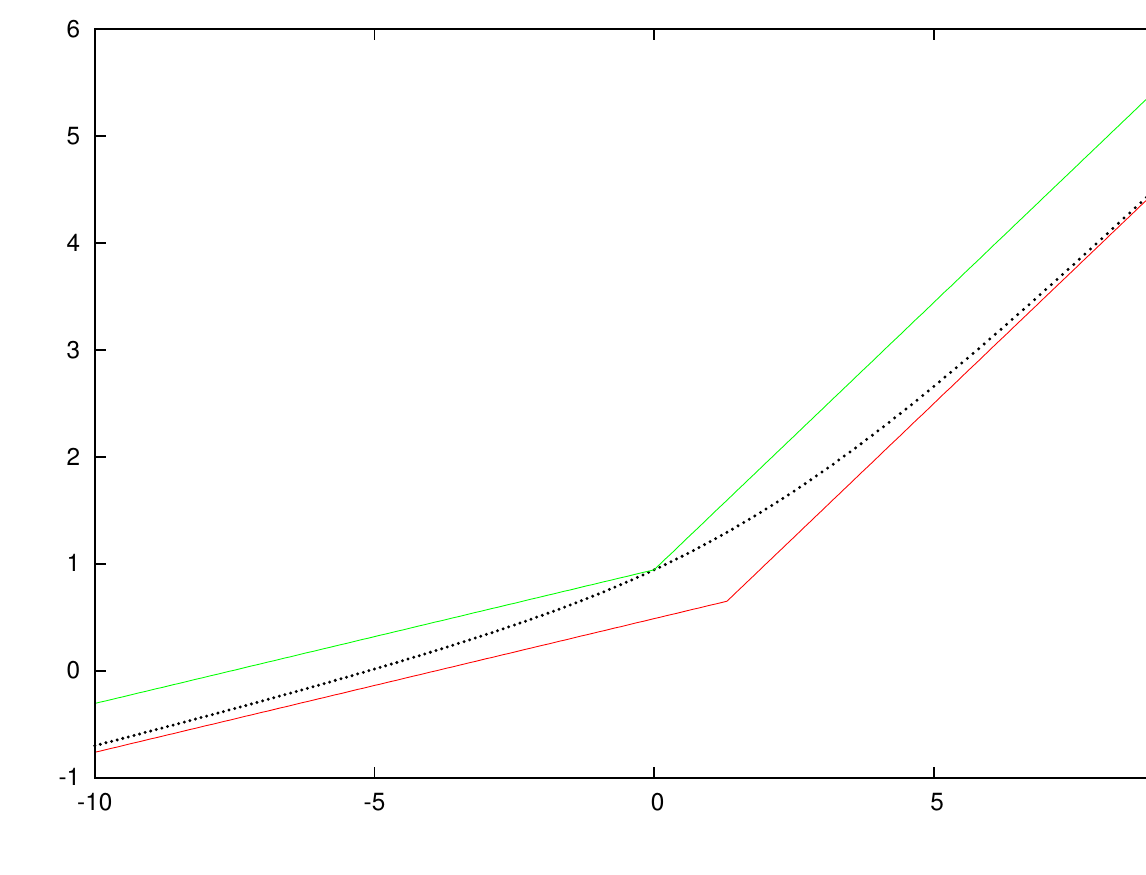}
\caption{Free energies in the $3\times1$ tube.}
\end{subfigure}
\vskip0.5cm
\begin{subfigure}{0.48\textwidth}
\includegraphics[width=\textwidth]{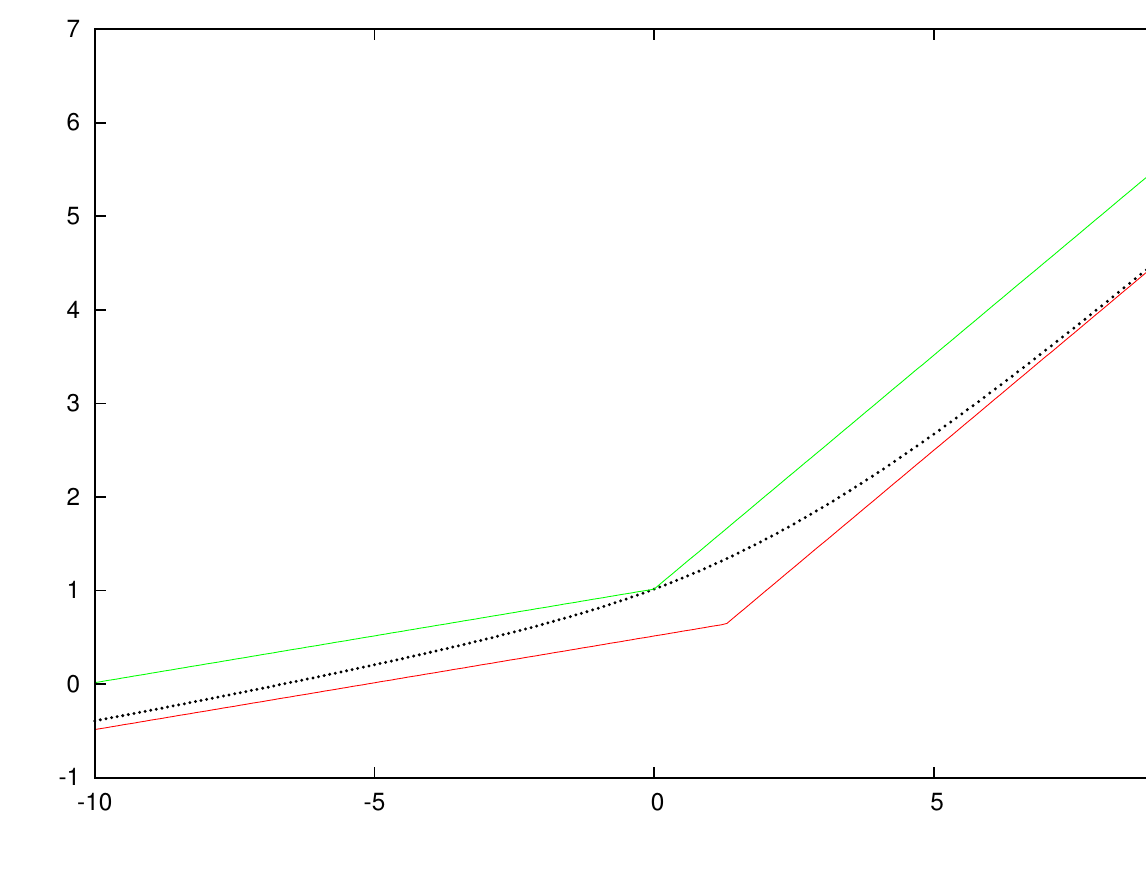}
\caption{Free energies in the $4\times1$ tube.}
\end{subfigure}
\hfill
\begin{subfigure}{0.48\textwidth}
\includegraphics[width=\textwidth]{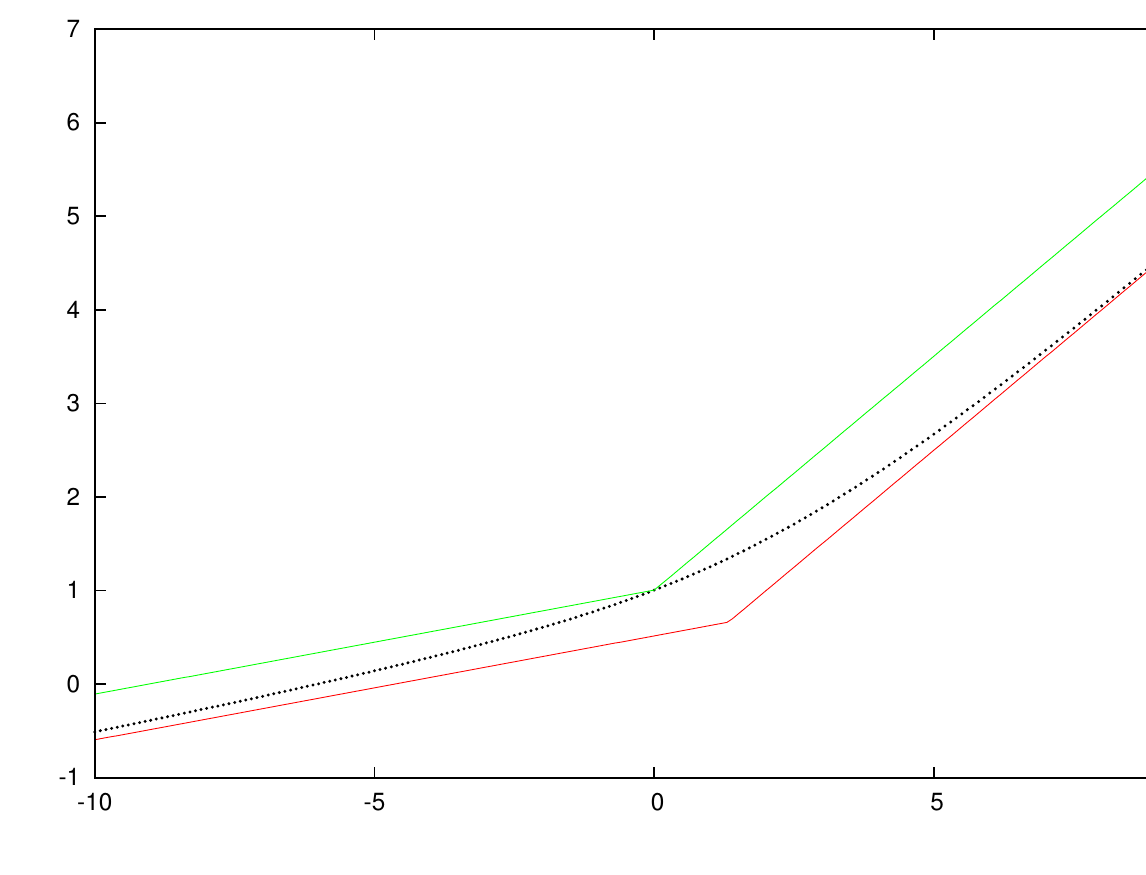}
\caption{Free energies in the $2\times2$ tube.}
\end{subfigure}
\caption{Numerical calculations of the free energies of polygons in three-dimensional tubes, plotted against the force $f$. The black points are calculations of $\mathcal F_\tube(f)$ (numerically accurate to $\pm 10^{-5}$). The red and green curves are respectively lower and upper bounds for $\mathcal F_\tube(f)$, as given by~\eqref{eqnfreeenergybounds}. Observe that in all cases, the black points appear to be asymptotic to the lower bounds for both $f\to\infty$ and $f\to-\infty$. }
\label{fig:fe_plots}
\end{figure}

\section{$f\to\pm\infty$ asymptotes}\label{sec:asymptotes}

In this section we focus on the free energy $\mathcal F_\tube(f)$. In particular, we 
determine its behaviour in the two large-force limits, $f\to\pm\infty$. There are a number of results from 
\cite[Chapter 3]{vanRensburg2000Statistical} (see also \cite[Chapter 3]{vanRensburg2015Statistical} {and \cite{2016arXiv160308553J} for modified presentations})
which will be important in this section. For this reason we explicitly state them here. We begin with some necessary assumptions.

\begin{assns}[Assumptions 3.1 of~\cite{vanRensburg2000Statistical}]\label{assns:31}
Let $u_k(m)$ be the number of objects of size $k$ and energy $m$. Assume that $u_k(m)$ satisfies the following properties:
\begin{itemize}
\item[(1)] There exists a constant $K>0$ such that $0\leq u_k(m) \leq K^k$ for each value of $k$ and $m$.
\item[(2)] {There exist  finite integers $A_k$ and $B_k$ and a real constant $C$  satisfying $0\leq A_k \leq B_k \leq Ck$ such that $u_k(m)>0$ for $A_k \leq m \leq B_k$ and $u_k(m)=0$ otherwise.}
\item[(3)] $u_k(m)$ satisfies a supermultiplicative inequality of the type
{\begin{equation}\label{eqn:xn_supermult}
u_{k_1}(m_1) u_{k_2}(m_2) \leq u_{k_1+k_2}(m_1+m_2).
\end{equation}}
\end{itemize}
\end{assns}

We now add a further assumption which is not required in~\cite{vanRensburg2000Statistical}, but will make calculations here somewhat simpler.

\begin{assns}\label{assns:A_and_B}
The limits
\[A = \lim_{k\to\infty} \frac{A_k}{k} \qquad\text{and}\qquad B = \lim_{k\to\infty} \frac{B_k}{k}\]
exist, with $A<B$.
\end{assns}

\begin{thm}[Theorems 3.4 and 3.5 of~\cite{vanRensburg2000Statistical}]\label{thm:34_35}
Let $u_k(m)$ be a sequence satisfying Assumptions~\ref{assns:31} and~\ref{assns:A_and_B}. Then if $\epsilon\in(A,B)$, the density function $\mathcal D(\epsilon)$ is defined by the limit
\[\log \mathcal D(\epsilon) = \lim_{k\to\infty}\frac{1}{k}\log u_k\left(\lfloor\epsilon k\rfloor\right).\]
The function $\log\mathcal D(\epsilon)$ is a concave function of $\epsilon$ on $(A,B)$, and is thus continuous and almost-everywhere differentiable. Moreover, there exists a number $\eta_k\in\{0,1\}$ such that for each $k$, 
\[\frac1k\log u_k\left(\lfloor\epsilon k\rfloor + \eta_k\right) \leq \log \mathcal D(\epsilon).\]
\end{thm}

We next define partition functions and relate them to the density function $\mathcal D(\epsilon)$. Let
\[U_k(z) = \sum_m u_k(m) e^{zm}.\]

\begin{thm}[{Theorems 3.6,} 3.17 and 3.19 of~\cite{vanRensburg2000Statistical}]\label{thm:317_319}
The limit
\[\mathcal F(z) = \lim_{k\to\infty}\frac1k \log U_k(z)\]
exists for all $z$. Moreover,
\[\mathcal F(z) = \sup_{A< \epsilon< B}\left\{\log \mathcal D(\epsilon) + \epsilon z\right\}\]
and
\[\log\mathcal D(\epsilon) = \inf_{-\infty<z<\infty}\left\{\mathcal F(z) - \epsilon z\right\}.\]
\end{thm}

Our next preliminary result is a generalisation of \cite{vanRensburg2000Statistical}*{equation (3.4)}. 

\begin{lem}\label{lem:Tn_lowerbound}
Let $T_k$ be a sequence satisfying $A_k\leq T_k\leq B_k$ and $T_k = Bk + o(k)$. Moreover, assume that $B_k < Bk$ for all $k$ sufficiently large. Then
\[{\log \mathcal D(B^-)\equiv} \lim_{\epsilon\to B^-} \log \mathcal D(\epsilon)  \geq \limsup_{k\to\infty}\frac1k \log u_k(T_k).\]
\end{lem}

\begin{proof}
Define $\epsilon_k = T_k/k$. Then because $T_k \leq B_k < Bk$, we have $\epsilon_k < B$ for all $k$ sufficiently large and $\lim_{k\to \infty} \epsilon_k=B$.

Fix any $k$ such that $\epsilon_k < B$.  Let $N\in \mathbb N$, and put $r=Nk$. Since $\epsilon_k r$ is an integer,  the supermultiplicativity assumption~\eqref{eqn:xn_supermult} can be used  repeatedly to split up $u_r(\epsilon_k r)$ a total of $N-1$ times, to obtain
\[u_r(\epsilon_k r) \geq u_k(\epsilon_k k)^N= u_k(T_k)^N.\]
Take logs, divide by $r=Nk$, and take $N \to \infty$ (keeping $k$ fixed). The limit of the left-hand-side exists, and is the log of the density function, so
\[\log \mathcal D(\epsilon_k) \geq \frac1k \log u_k(T_k).\]
Taking the $\limsup$ as $k\to\infty$ of both sides then gives
\begin{align*}
\limsup_{k\to\infty} \frac1k \log u_k(T_k) &\leq \limsup_{k\to\infty} \log \mathcal D(\epsilon_k) \\
&\leq \lim_{\epsilon\to B^-} \log \mathcal D(\epsilon) \\
&= \log \mathcal D(B^-),
\end{align*}
where the final limit exists due to the concavity of $\log\mathcal D(\epsilon)$.
\end{proof}

We also note the following {consequences} of the concavity of $\log \mathcal D(\epsilon)$ and Theorem~\ref{thm:317_319} (see for example \cite[Corollary 4]{Madras_1988} and \cite[Chapter VI]{Ellis_1985} for further background on convex functions and Legendre transforms):
{

\begin{align}
\mathcal \displaystyle{\lim_{z\to\infty} (F(z) -Bz)}&=  \lim_{\epsilon\to B^-} \log \mathcal D(\epsilon) \equiv  \log \mathcal D(B^-) \label{eqn:sup_z_infty} \\
\mathcal \displaystyle{\lim_{z\to -\infty} (F(z)-Az)} &=  \lim_{\epsilon\to A^+} \log \mathcal D(\epsilon)\equiv  \log \mathcal D(A^+)  .\label{eqn:sup_z_minus_infty}
\end{align}

}

\subsection{$f\to\infty$}\label{ssec:infinity_asymp}

The main result of this section is the following theorem.

\begin{thm}\label{thm:f_infinity}
For any tube size $L\times M$, in the limit $f\to\infty$ the free energy $\mathcal F_\tube(f)$ is asymptotic to $f/2$. That is,
\begin{equation}\label{eqn:f_infinity}
\lim_{f\to\infty}\left(\mathcal F_\tube(f) - \frac{f}{2}\right) = 0.
\end{equation}
\end{thm}

Theorem~\ref{thm:f_infinity} is in fact a corollary of a more general result. We restrict polygons to the half-space of $\mathbb Z^3$ defined by $x\geq 0$. Let $\mathcal P$ be the {subset of these} polygons which contain at least one edge in the plane $x=0$;
the number of such polygons (counted up to $y$- and $z$-translations) is equal to $p_n$ as previously defined in Section \ref{sec:themodel}.
The span of these polygons is defined in the same way as for those in $\tube$; let $p_n(s)$ be the number with length $n$ and span $s$, and define the partition function
\[Z_n(f) = \sum_{s\geq 0}p_n(s)e^{fs}.\]
It is well-known~\cite{vanRensburgetal2008stretched} that the free energy
\[\mathcal F(f) = \lim_{n\to\infty}\frac1n \log Z_n(f)\]
exists for all $f$ and is a convex function.

\begin{thm}\label{thm:allpolys_f_infinity}
In the limit $f\to\infty$, the free energy $\mathcal F(f)$ is asymptotic to $f/2$. That is,
\begin{equation}\label{eqn:allpolys_f_infinity}
\lim_{f\to\infty}\left(\mathcal F(f) - \frac{f}{2}\right) = 0.
\end{equation}
\end{thm}

Before commencing the proof, we introduce some new definitions. Let $\mathcal P^*$ be the set of polygons $\pi\in\mathcal P$ which satisfy the additional constraints:
\begin{itemize}
\item $\pi$ has span $s\geq 2$,
\item $\pi$ contains the edge $(0,0,0)\edge(0,1,0)$ {(called its left-most-edge)} and no other edges in the plane $x=0$,
\item $\pi$ contains the edge $(s,y,z)\edge(s,y+1,z)$ for some $y$ and $z$ {(called its right-most-edge)}, and contains no other edges in the plane $x=s$, and
\item $\pi$ contains no edges in the plane $x=s-1$.
\end{itemize}
Let $p^*_n(s)$ be the number of polygons in $\mathcal P^*$ with length $n$ and span $s$. Then $p^*_n(s)$ satisfies Assumptions~\ref{assns:31}, with length corresponding to size and span corresponding to energy. To see this, note the following.
\begin{itemize}
\item[(1)] {$K=6$} satisfies condition (1).
\item[(2)] The numbers $A_n$ and $B_n$ are
\[A_n = \begin{cases} 2 & n=6 \\ 3 & n=8 \\ 4 & n\geq 10 \end{cases} \qquad\qquad B_n = \frac{n-2}{2}.\]
{The $n$-edge polygon $\tilde{\pi}_n\in \mathcal P^*$ consisting of the edges $(0,0,0)-(0,1,0), (\frac{n-2}{2},0,0)-(\frac{n-2}{2},1,0)$ and  $(i,1,0)-(i+1,1,0), (i,0,0)-(i+1,0,0), i=0,..., \frac{n-2}{2}-1$ has span $B_n$. Note that $A_n=B_n$ for $n\leq 8$. For $n\geq 10$,  an $n$-edge polygon in $\mathcal P^*$ with span $s\in[A_n,B_n)$
can be obtained from $\tilde{\pi}_{2s+2}$  by concatenating an appropriately rotated and translated version of  $\tilde{\pi}_{n-2s-2}$ at the edge $(1,1,0)-(2,1,0)$ of $\tilde{\pi}_{2s+2}$.  Thus $p^*_n(s)>0$.}

\item[(3)] Any two polygons  $\pi_1,\pi_2\in \mathcal P^*$ can be concatenated (by translating $\pi_2$ so that its left-most-edge coincides with the right-most-edge of $\pi_1$ and then deleting the two coincident edges) in a way that preserves total length and total span, giving
\[p^*_{n_1}(s_1)p^*_{n_2}(s_2) \leq p_{n_1+n_2}(s_1+s_2).\]
\end{itemize}
Now define $P^*_n(f) = \sum_s p^*_n(s)e^{fs}$. By Theorem~\ref{thm:317_319}, the free energy
\[\mathcal F^*(f) = \lim_{n\to\infty}\frac1n \log P^*_n(f)\]
exists. Since $p^*_n(s) \leq p_n(s)$, we 
have $\mathcal F^*(f) \leq \mathcal F(f)$. Moreover, there exist constants $n_0$ and $s_0$ such that any polygon $\pi\in\mathcal P$ of length $n$ and span $s$ can be converted into a unique polygon $\pi'\in\mathcal P^*$ with length $n+n_0$ and span $s+s_0$. So
\[p_n(s) \leq p^*_{n+n_0}(s+s_0).\]
Multiply this by $e^{f(s+s_0)}$, sum over $s$, take logs, divide by $n$ and take $n\to\infty$ to obtain $\mathcal F(f) \leq \mathcal F^*(f)$, so that we in fact have 
\begin{equation}\label{eqn:hp_F=F*}
\mathcal F^*(f) = \mathcal F(f).
\end{equation}

\begin{proof}[Proof of Theorem~\ref{thm:allpolys_f_infinity}]

By Theorems~\ref{thm:34_35} and~\ref{thm:317_319}, the Legendre transform of $\mathcal F^*$,
\begin{equation}\label{eqn:legendre_of_F}
\log\mathcal S^*(\epsilon) = \inf_{-\infty<f<\infty}\left\{\mathcal F^*(f) - \epsilon f\right\}= \lim_{n\to\infty}\frac1n \log p^*_n\left(\lfloor\epsilon n\rfloor\right),
\end{equation}
exists and is finite and concave for $\epsilon\in(0, 1/2)$,
where $\mathcal S^*(\epsilon)$ can be viewed as the growth rate of polygons with ``span density'' $\epsilon$, that is, those polygons whose span is asymptotically $\epsilon$ times their length.

Then by Theorem~\ref{thm:317_319},
\begin{equation}\label{eqn:F_S_inverse_legendre}
\mathcal F^*(f) = \sup_{0<\epsilon<1/2}\left\{\log\mathcal S^*(\epsilon) + f\epsilon\right\}.
\end{equation}
Then as $f$ gets large, it follows from~\eqref{eqn:sup_z_infty} {that the behaviour of $\mathcal F^*(f)$}
is obtained by taking $\epsilon\to (1/2)^-$. 
We thus need to examine the behaviour of $\log \mathcal S^*(\epsilon)$ in this limit.

First note that by applying Lemma~\ref{lem:Tn_lowerbound} with the sequence $T_n=(n-2)/2$, we have
\begin{equation}\label{eqn:S_epsilon_nonneg}
\lim_{\epsilon\to1/2^-}\log\mathcal S^*(\epsilon) \geq \limsup_{n\to\infty}\frac1n\log p^*_n\left(\frac{n-2}{2}\right) = 0.
\end{equation}

Now polygons in $\mathcal P^*$ can be unambiguously rooted and oriented (let $(0,0,0)$ be the root, with the first step in {the} positive $y$ direction), {so we can view such a polygon as a walk  which is self-avoiding except for the start and end vertex.  Given $\pi\in \mathcal P^*_n$, let $\omega(\pi)$ be the resulting walk composed of the sequence of vertices $v_0=(0,0,0), v_1, ...., v_n, v_{n+1}=v_0$.   We define an \emph{increasing step} of $\pi$ to be any step $(v_i,v_{i+1})$ of $\omega(\pi)$  in the positive $x$ direction which increases the span of the walk (i.e. the maximum $x$-coordinate of the vertices in the subwalk from $v_0$ to $v_{i+1}$ is one greater than that for the subwalk from $v_0$ to $v_{i}$). So a polygon with span $s$ has  exactly $s$ increasing steps. Likewise, define the \emph{decreasing} steps of $\pi$ to be the increasing steps of $\omega(\pi)'$, where $\omega(\pi)'$ is the walk obtained by reversing the orientation of $\omega(\pi)$ (but maintaining the same root). A polygon of span $s$ will thus also have $s$ decreasing steps.}

To obtain an upper bound on $\log\mathcal S^*(\epsilon)$ as $\epsilon\to1/2^-$, we define
\[k^*_n(t) = \sum_{s \geq t} p^*_n(s),\]
that is, the number of polygons of length $n$ and span at least $t$. 

Given any fixed $r\leq n$, we can write $n=pr+q$ with $0\leq q<r$, so any polygon $\pi\in \mathcal P^*_n$ can be divided into $p$ or $p+1$ subwalks, the first $p$ of which have length $r$.
If the polygon's span is at least $t$ then it has  at least $t$ increasing and at least $t$ decreasing steps, and thus at most $n-2t$ steps which are neither increasing nor decreasing. So at most $n-2t$ of its $p$ length-$r$ subwalks contain non-increasing or non-decreasing steps, and the rest (for $p>n-2t$) must be composed entirely of increasing or decreasing steps. A subwalk that contains only increasing or decreasing steps must only have steps in the $x$ direction (positive or negative),  and hence (due to self-avoidance) the subwalk must be either entirely increasing or entirely decreasing. Hence there are only two types of such subwalks of length $r$; one consists of $r$ positive $x$-steps and the other $r$ negative $x$-steps. Letting $u=n-2t$, we thus have
\begin{equation}\label{eqn:cut_up_polygon}
k^*_n(t) \leq \sum_{i=0}^{u}\binom{p}{i}c_r^i 2^{p-i}c_q,
\end{equation}
where $c_n$ is the number of SAWs of length $n$.

Given any $\delta>0$, take $r$ sufficiently large ($\geq N_{\delta}$) so that $2\leq e^{\delta r}$ and $c_r\leq e^{(\delta+\kappa)r}$ (this is possible due to \eqref{eqnmudef}). 
Then
\begin{align}
k^*_n(t) &\leq \sum_{i=0}^{u}\binom{p}{i}\left(e^{(\delta+\kappa)r}\right)^i\left(e^{\delta r}\right)^{p-i}c_q \notag\\
&= e^{\delta r p}c_q\sum_{i=0}^{u}\binom{p}{i}e^{\kappa r i}.\label{eqn:ks_n_bound}
\end{align}

Let $t=\lfloor \epsilon n\rfloor$ so that $u=n-2\lfloor \epsilon n\rfloor$. Noting that $p\sim n/r$, let $\epsilon$ be sufficiently close to $1/2$ so that $u<p/2$ {(for $p\geq 4$, $\epsilon > (1/2)-1/(3r)$ is sufficient)}. Then the largest summand of~\eqref{eqn:ks_n_bound} is the last one, so

\[k^*_n(\lfloor \epsilon n\rfloor) \leq e^{\delta r p}c_q(u+1)\binom{p}{u}e^{\kappa r u}.\]

Take logs, divide by $n$ and apply Stirling's {formula}:
\begin{multline*}
\frac1n \log k^*_n\left(\lfloor \epsilon n\rfloor\right) \leq \frac1n \log\left(u+1\right) + \frac{\delta r p}{n} + \frac1n \log c_q + \frac{\kappa ru}{n} \\ - \frac{p}{n}\log\left(\frac{p-u}{p}\right) + \frac{u}{n}\log\left(\frac{p-u}{u}\right) + O\left(\frac{\log n}{n}\right).
\end{multline*}
Then for $r>N_{\delta}$ and $\epsilon>(1/2)-1/(3r)$ fixed,  take $p\to\infty$ and hence $n\to\infty$ (note that $u\sim (1-2\epsilon)n$):
\begin{align*}
\log \mathcal S^*(\epsilon) &\leq \limsup_{n\to\infty}\frac1n\log k^*_n\left(\lfloor\epsilon n\rfloor\right) \\
&\leq \delta +\kappa r(1-2\epsilon) - \frac1r \log\left(1-r+2r\epsilon\right) +(1-2\epsilon)\log\left(\frac{1-r+2r\epsilon}{r-2r\epsilon}\right).
\end{align*}
Taking $\epsilon \to 1/2^-$ gives
\[\limsup_{\epsilon\to1/2^-} \log \mathcal S^*(\epsilon) \leq \delta.\]
Let $\delta$ be arbitrarily small, and combine with~\eqref{eqn:S_epsilon_nonneg}, to obtain
\begin{equation}\label{eqn:Sepsilon_to0}
\lim_{\epsilon\to1/2^-}\log\mathcal S^*(\epsilon) = 0.
\end{equation}

Finally by taking $f\to\infty$ in~\eqref{eqn:F_S_inverse_legendre}, and using~\eqref{eqn:sup_z_infty} and~\eqref{eqn:hp_F=F*}, we obtain the result.
\end{proof}

The corresponding result for polygons in $\tube$ then follows in a straightforward manner {as described next}.
\begin{proof}[Proof of Theorem~\ref{thm:f_infinity}]
Since $\mathcal P_{\tube,n}$ contains at least one polygon of span $(n-2)/2$ for every even $n$ (specifically $\tilde{\pi}_n$), we have $\mathcal F_\tube(f) \geq f/2$. 

Every polygon in $\tube$ also occurs in the half-space, but certain polygons which are only counted once in $Z_n(f)$ may be counted multiple times in $Z_{\tube,n}(f)$, because translations of a polygon in the $y$ and/or $z$ directions (but still staying in $\tube$) are all counted separately. However, the number of possible translations is bounded above by a constant $c$ depending only on $L$ and $M$, so
\[Z_{\tube,n}(f) \leq cZ_n(f).\]
Taking logs, dividing by $n$ and sending $n\to\infty$, we have
\[\mathcal F_\tube(f) \leq \mathcal F(f),\]
and the result follows.
\end{proof}

\subsection{$f\to-\infty$}\label{ssec:f_minus_infinity}

In this section we consider the case of compressed polygons. Some preliminary definitions and results are required before the main theorem can be stated.

Given a polygon $\pi\in\tube$, a \emph{hinge} $H_k$ of $\pi$ is the set of edges and vertices lying in the intersection of $\pi$ and the $y$-$z$ plane defined by $\{(x,y,z):x=k\}$. A \emph{section} $S_k$ is the set of edges in $\pi$, in the $x$ direction, connecting $H_{k-1}$ and $H_k$. A \emph{half-section} of $S_k$ is the set of half-edges in $S_k$ with either $k-1\leq x\leq k-\frac12$ or $k-\frac12\leq x\leq k$.

A \emph{1-block} of $\tube$ is any non-empty hinge which can occur in a polygon $\pi$ in $\tube$, together with the half-edges of $\pi$ in the two adjacent half-sections. The \emph{length} of a 1-block is the sum of the lengths of all its {polygon} edges and half-edges. It is thus natural to view a 1-block as the part of a polygon between two half-integer $x$-coordinates $k\pm\frac12$ for some $k\in\mathbb Z$.

An \emph{$s$-block} is then any 
{connected sequence}
of $s$ 1-blocks, the entirety of which can occur in a polygon in $\tube$. (It is also possible, if the first and last half-sections of the $s$-block are empty, for the $s$-block itself to be a polygon.) The length of an $s$-block is the sum of the lengths of its constituent 1-blocks. Let $b_{\tube,s}$ be the number of $s$-blocks in $\tube$, counted up to translation in the $x$-direction. See Figure~\ref{fig:sblock} for an example of a 9-block in a $6\times0$ tube.

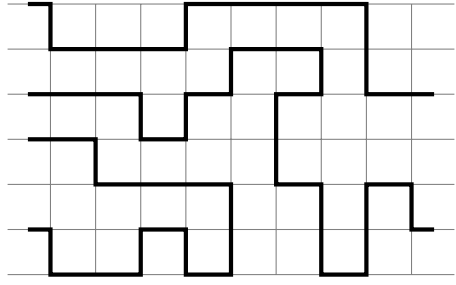
\begin{figure}
\centering
\begin{tikzpicture}[scale=0.6]
\draw[step = 1cm, gray, thin] (0.05,0) grid (9.95,6);
\draw[ultra thick] (0.5,6) -- (1,6) -- (1,5) -- (4,5) -- (4,6) -- (8,6) -- (8,4) -- (9.5,4);
\draw[ultra thick] (0.5,4) -- (3,4) -- (3,3) -- (4,3) -- (4,4) -- (5,4) -- (5,5) -- (7,5) -- (7,4) -- (6,4) -- (6,2) -- (7,2) -- (7,0) -- (8,0) -- (8,2) -- (9,2) -- (9,1) -- (9.5,1);
\draw[ultra thick] (0.5,3) -- (2,3) -- (2,2) -- (5,2) -- (5,0) -- (4,0) -- (4,1) -- (3,1) -- (3,0) -- (1,0) -- (1,1) -- (0.5,1);
\end{tikzpicture}
\caption{A 9-block of the $6\times0$ tube. This 9-block has length 50.}
\label{fig:sblock}
\end{figure}

\begin{lem}\label{lem:sblocks_growthrate}
The limit
\begin{equation}\label{eqn:sblocks_growthrate}
\beta_\tube = \lim_{s\to\infty} \frac1s \log b_{\tube,s}
\end{equation}
exists and is finite.
\end{lem}

\begin{proof}
Any $(s+t)$-block can be cut into an $s$-block and a $t$-block; we thus have
\[b_{\tube,s+t} \leq b_{\tube,s}b_{\tube,t}.\]
So $\{\log b_{\tube,s}\}$ is a subadditive sequence, and the limit~\eqref{eqn:sblocks_growthrate} exists. We clearly have $b_{\tube,s}\geq1$ for all $s\geq 1$, so that $\beta_\tube$ is finite.
\end{proof}

A 1-block is \emph{full} if its length is equal to $W=(L+1)(M+1)$. Equivalently, a 1-block is full if {every vertex in a plane $\{(x,y,z):x=k\}$ is in its hinge}. An $s$-block is full if every one of its constituent 1-blocks is full. Let $\bfull_{\tube,s}$ be the number of full $s$-blocks in $\tube$.

\begin{lem}\label{lem:full_sblocks_growthrate}
The limit
\begin{equation}\label{eqn:full_sblocks_growthrate}
\betafull_\tube = \lim_{s\to\infty} \frac1s \log \bfull_{\tube,s}
\end{equation}
exists and is finite.
\end{lem}

\begin{proof}
The reasoning is the same as in Lemma~\ref{lem:sblocks_growthrate}. A full $(s+t)$-block can be cut into a full $s$-block and a full $t$-block, so
\[\bfull_{\tube,s+t} \leq \bfull_{\tube,s}\bfull_{\tube,t}.\]
The sequence $\{\log \bfull_{\tube,s}\}$ is thus subadditive, and the limit~\eqref{eqn:full_sblocks_growthrate} exists. Likewise {(consider for example $s$-blocks obtained from Hamiltonian polygons)} $\bfull_{\tube,s}\geq 1$ for all $s\geq1$.
\end{proof}

We are now able to state the main theorem of this section.

\begin{thm}\label{thm:f_minus_infinity}
For any tube size $L\times M$, in the limit $f\to-\infty$ the free energy $\mathcal F_\tube(f)$ is asymptotic to $(\betafull_\tube+f)/W$, where $W=(L+1)(M+1)$. That is,
\begin{equation}\label{eqn:f_minus_infinity}
\lim_{f\to-\infty}\left(\mathcal F_\tube(f) - \frac{f}{W}\right) = \frac{\betafull_\tube}{W}.
\end{equation}
\end{thm}

The proof of Theorem~\ref{thm:f_minus_infinity} will require, at least at first, a different approach to that of Theorem~\ref{thm:f_infinity}. We begin with some more definitions.

Let $\mathcal P^*_\tube$ be the set of those polygons $\pi\in\mathcal P_\tube$ which satisfy the additional constraints:
\begin{itemize}
\item $\pi$ has span $s\geq 2$,
\item $\pi$ contains the edge $(0,0,0)\edge(0,1,0)$ and no other edges in the plane $x=0$,
\item $\pi$ contains the edge $(s,0,0)\edge(s,1,0)$ and no other edges in the plane $x=s$, and
\item $\pi$ contains no edges in the plane $x=s-1$.
\end{itemize}
Let $p^*_{\tube,n}(s)$ be the number of polygons in $\mathcal P^*_{\tube}$ with length $n$ and span $s$. We define a partition function analogous to $Z_{\tube,n}(f)$:
\[Z^*_{\tube,n}(f) = \sum_s p^*_{\tube,n}(s) e^{fs}.\]

\begin{lem}\label{lem:F*_same_as_F}
The free energy 
\[\mathcal F^*_\tube(f) = \lim_{n\to\infty}\frac1n \log Z^*_{\tube,n}(f)\]
exists and is equal to $\mathcal F_\tube(f)$.
\end{lem}

\begin{proof}
If $(L,M)=(1,0)$ then $Z^*_{\tube,n}(f) = e^{f(n-2)/2}$, and the result is trivial. Otherwise, at least one of the statements $L\geq 2$ or $M\geq 1$ is true.

We show that the sequence $p^*_{\tube,n}(s)$ satisfies Assumptions~\ref{assns:31} with size $k=n$ and energy $m=s$, so that Theorem~\ref{thm:317_319} can be applied.
\begin{enumerate}
\item Using $K=6$ suffices to satisfy condition (1).
\item The numbers $A_n$ and $B_n$ (respectively the minimum and maximum possible spans for a $\mathcal P^*_\tube$ polygon of length $n$) are
{\[A_n = \begin{cases} 2 & n=6 \\ 3 & n=8 \\ \max\left\{ 4, \left\lceil\frac{n-6}{W}\right\rceil + 2 \right\} & n\geq 10 \end{cases} \qquad\qquad B_n = \frac{n-2}{2}.\]
Using specific hinges such as those defined in Section  \ref{sec:hamiltonian}  for the proof of Theorem \ref{thm:ham_growthrate}, it is possible to prove that $p^*_{\tube,n}(s)>0$ for each integer $s\in [A_n,B_n]$.}
\item The set $\mathcal P^*_\tube$ has been defined so that any two polygons $\pi_1,\pi_2$ in $\mathcal P^*_\tube$ can be concatenated in a way that preserves both total length and total span. Let $\pi_1$ have span $s_1$, and define $e_1$ to be the single edge of $\pi_1$ with maximal $x$-coordinate and $e_2$ to be the single edge of $\pi_2$ with minimal $x$-coordinate. Then
\begin{enumerate}
\item[i.] Translate $\pi_2$ so that $e_1$ and $e_2$ coincide, and delete those two edges.
\item[ii.] If $L\geq 2$ then replace the edge $(s_1-1,1,0)\edge(s_1,1,0)$ with the three edges
\[(s_1-1,1,0)\edge(s_1-1,2,0)\edge(s_1,2,0)\edge(s_1,1,0).\]
Otherwise if $(L,M)=(1,1)$ then replace the edge $(s_1-1,1,0)\edge(s_1,1,0)$ with the three edges
\[(s_1-1,1,0)\edge(s_1-1,1,1)\edge(s_1,1,1)\edge(s_1,1,0).\]
\end{enumerate}
See Figure~\ref{fig:concatenation} for an illustration. So any two polygons $\pi_1,\pi_2$ in $\mathcal P^*_\tube$, of lengths $n_1$ and $n_2$ and spans $s_1$ and $s_2$, can be concatenated to give another polygon in $\mathcal P^*_\tube$ of length $n_1+n_2$ and span $s_1+s_2$. Thus
\begin{equation}\label{eqn:ps_tube_submult}
p^*_{\tube,n_1}(s_1)p^*_{\tube,n_2}(s_2) \leq p^*_{\tube,n_1+n_2}(s_1+s_2),
\end{equation}
and condition (3) is satisfied.
\end{enumerate}

\begin{figure}
\centering
%\begin{tikzpicture}
%\tikzset{scale=0.5, every node/.style={circle, fill, inner sep=2.5pt}}
%\begin{knot}[clip width=3]
%\strand[ultra thick] (0,0) node{} -- (1,1) node{} -- (4,1) node{} -- (5,2) node{} -- (5,5) node{} -- (8,5) node{} -- (8,2) node{} -- (7,1) node{} -- (10,1) node{};
%\draw (13,1) node{};
%\begin{scope}[on background layer] 
%\draw[ultra thick, red] (13,1) -- (12,0);
%\draw[ultra thick, green] (10,1) -- (13,1);
%\end{scope}
%\strand[ultra thick] (12,0) node{} -- (9,0) node{} -- (6,0) node{} -- (6,3) node{} -- (7,4) node{};
%\strand[ultra thick] (7,4) node{} -- (4,4) node{} -- (3,3) node{} -- (3,0) node{} -- (0,0) node{};
%\scoped[on background layer] \draw[ultra thick, red] (18,0) -- (19,1);
%\strand[ultra thick] (19,1) node{} -- (22,1) node{} -- (22,4) node{} -- (23,5) node{} -- (23,2) node{} -- (26,2) node{} -- (25,1) node{} -- (28,1) node{} -- (31,1) node{} -- (30,0) node{} -- (27,0) node{} -- (24,0) node{} -- (21,0) node{} -- (18,0) node{};
%\flipcrossings{1,2}
%\end{knot}
%\tikzset{every node/.style={}}
%\draw (15.5,2.5) node{\LARGE $+$};
%\end{tikzpicture}
\includegraphics{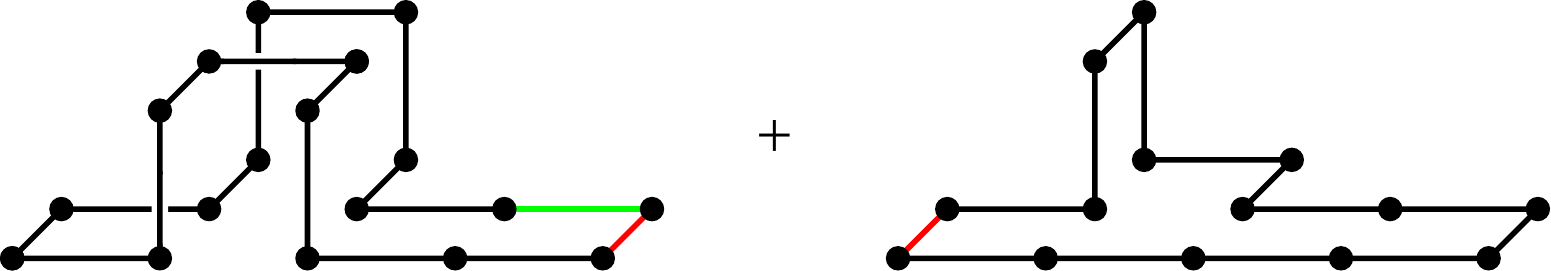}
\vskip1cm
%\begin{tikzpicture}
%\tikzset{scale=0.5}
%\draw(0,2.5) node {\LARGE $=$};
%\tikzset{every node/.style={circle, fill, inner sep=2.5pt}}
%\begin{knot}[clip width=3]
%\strand[ultra thick] (1,0) node{} -- (2,1) node{} -- (5,1) node{} -- (6,2) node{} -- (6,5) node{} -- (9,5) node{} -- (9,2) node{} -- (8,1) node{} -- (11,1) node{};
%\begin{scope}[on background layer]
%\draw[ultra thick, blue] (11,1) -- (12,2) -- (15,2) -- (14,1);
%\end{scope}
%\draw (12,2) node{};
%\draw (15,2) node{};
%\strand[ultra thick] (14,1) node{} -- (17,1) node{} -- (17,4) node{} -- (18,5) node{} -- (18,2) node{} -- (21,2) node{} -- (20,1) node{} -- (23,1) node{} -- (26,1) node{} -- (25,0) node{} -- (22,0) node{} -- (19,0) node{} -- (16,0) node{} -- (13,0) node{} -- (10,0) node{} -- (7,0) node{} -- (7,3) node{} -- (8,4) node{} -- (5,4) node{} -- (4,3) node{} -- (4,0) node{} -- (1,0) node{};
%\flipcrossings{1,2}
%\end{knot}
%\end{tikzpicture}
\includegraphics{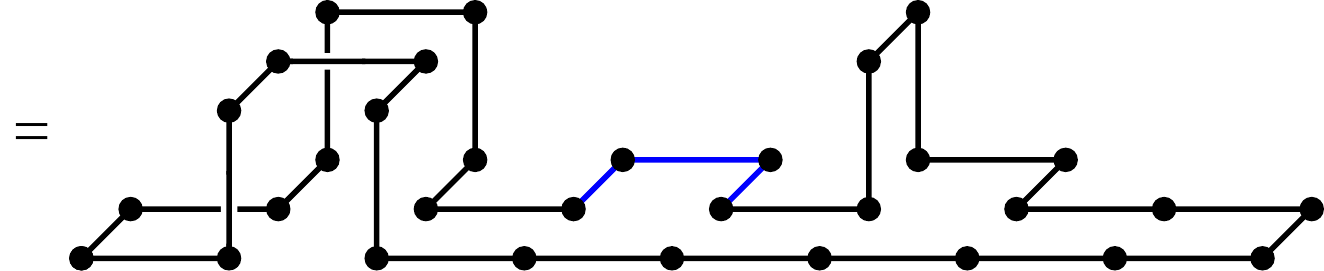}
\caption{The concatenation operation of $\mathcal P^*_\tube$ polygons described in the proof of Lemma~\ref{lem:F*_same_as_F}, in the $2\times1$ tube. The second polygon is translated so that the red edges coincide. These edges are then removed, and the green edge is replaced by the three blue edges. Note that the total length{, 32, and the total span, 8,} are preserved.}
\label{fig:concatenation}
\end{figure}

Since $\mathcal P^*_\tube \subseteq \mathcal P_\tube$, we have $\mathcal F^*_\tube(f) \leq \mathcal F_\tube(f)$. To obtain the reverse inequality, we use the fact that any $\mathcal P_\tube$ polygon can be converted into a unique $\mathcal P^*_\tube$ polygon by adding a fixed number $n_0$ of edges, which increase the span by at most a constant number $s_0$ {(see for example \cite{Sot89,Atapour2008Thesis})}. (Both $n_0$ and $s_0$ depend on the dimensions of the tube $\tube$.) Thus
\[p_{\tube,n}(s) \leq \sum_{s'=s}^{s+s_0}p^*_{\tube,n+n_0}(s').\]
Multiplying by $e^{fs}$ and summing over $s$,
\begin{align*}
Z_n(f) = \sum_s p_{\tube,n}(s)e^{fs} &\leq \sum_s e^{fs} \sum_{s'=s}^{s+s_0}p^*_{\tube,n+n_0}(s') \\
&= \sum_s \sum_{s'=s}^{s+s_0} p^*_{\tube,n+n_0}(s') e^{fs'}e^{f(s-s')} \\
&= \left(1+e^{-f} + \ldots + e^{-fs_0}\right)\sum_s p^*_{\tube,n+n_0}(s) e^{fs} \\
&\leq (s_0+1)\max\{1,e^{-fs_0}\} Z^*_{\tube,n+n_0}(f).
\end{align*}
Taking logs, dividing by $n$ and letting $n\to\infty$ provides the required result.
\end{proof}

Polygons in $\mathcal P^*_\tube$ then have a density function, similar to $\mathcal S^*(\epsilon)$ as defined in~\eqref{eqn:legendre_of_F}:
\[\log \mathcal S^*_\tube(\epsilon) = \inf_{-\infty<f<\infty} \left\{\mathcal F^*_\tube(f) - \epsilon f\right\} = \lim_{n\to\infty} \frac1n \log p^*_{\tube,n}(\lfloor \epsilon n\rfloor)\]
for $\epsilon \in (1/W,1/2)$, with
\begin{equation}\label{eqn:Fstube_inverse_Legendre}
\mathcal F^*_\tube(f) = \sup_{1/W<\epsilon<1/2}\left\{\log \mathcal S^*_\tube(\epsilon) + \epsilon f\right\}.
\end{equation}

The approach to proving Theorem~\ref{thm:f_minus_infinity} will involve the `dual' object to $\mathcal F^*_\tube(f)$. Let $q^*_{\tube,s}(n) = p^*_{\tube,n}(s)$. (We introduce this quantity to make it clear that we are now interpreting the span of a polygon as its `size' and the length of a polygon as its `energy'.)
Define
\[Q^*_{\tube,s}(z) = \sum_n q^*_{\tube,s}(n) e^{zn}.\]

\begin{lem}\label{lem:dual_fe}
The free energy
\[\mathcal G^*_\tube(z) = \lim_{s\to\infty}\frac1s \log Q^*_{\tube,s}(z)\]
exists for all $z$. It is a convex function of $z$, and is thus continuous and almost-everywhere differentiable.
\end{lem}

\begin{proof}
If $(L,M)=(1,0)$ then the result is again trivial, so we can assume that at least one of the statements $L\geq2$ or $M\geq 1$ is true.

We show that the sequence $q^*_{\tube,s}(n)$ satisfies Assumptions~\ref{assns:31}, with one minor caveat.
\begin{itemize}

\item[(1)] Since {$q^*_{\tube,s}(n) \leq b_{\tube,s+1}\leq (b_{\tube,1})^{s+1}$}, using $K=(b_{\tube,1})^2$ suffices to satisfy condition (1).

\item[(2)] The numbers $A_s$ and $B_s$ (respectively the minimum and maximum possible lengths of a $\mathcal P^*_\tube$ polygon of span $s$) are
\[A_s = 2(s+1) \qquad \qquad B_s = \begin{cases} A_s & \text{if } s=2,3 \\ W(s-2) + 6 & \text{if } W \text{ or } s\geq 4 \text{ even} \\ W(s-2) + 5 & \text{if } W \text{ and } s> 4 \text{ odd}.\end{cases}\]
\noindent However, note that $q^*_{\tube,s}(n) > 0$ only if $n$ is even. Condition (2) can then be met by letting the energy of a polygon be its \emph{half-length}, rather than its length. Adjusting everything to account for this essentially amounts to taking $n\mapsto n/2$ in the definitions of $q^*_{\tube,s}(n)$ and $Q^*_{\tube,s}(z)$, and likewise dividing the values of $A_s$ and $B_s$ by 2. This is straightforward, so we will in general continue to use length instead of half-length.

\item[(3)] The inequality~\eqref{eqn:ps_tube_submult} can be rewritten as
\[q^*_{\tube,s_1}(n_1)q^*_{\tube,s_2}(n_2) \leq q^*_{\tube,s_1+s_2}(n_1+n_2),\]
so condition (3) is satisfied.
\end{itemize}

By Theorem~\ref{thm:317_319}, the free energy $\mathcal G^*_{\tube}(z)$ exists. A standard application of the Cauchy-Schwarz inequality (see for example~\cite[Section 2.3]{Hammersley1982Selfavoiding}) demonstrates the convexity of $\mathcal G^*_\tube(z)$.
\end{proof}

We will now determine the asymptotic behaviour of $\mathcal G^*_\tube(z)$ as $z\to\infty$, and will see later that this is related, in a very simple way, to the behaviour of $\mathcal F_\tube(f)$ as $f\to-\infty$. We once again make use of a density function. By Theorem~\ref{thm:317_319} there is a `length density' function, analogous to $\mathcal S^*(\epsilon)$ as defined in~\eqref{eqn:legendre_of_F}:
\begin{equation}\label{eqn:length_density}
\log \mathcal L^*_\tube(\alpha) = \inf_{-\infty<z<\infty}\left\{\mathcal G^*_\tube(z)- \alpha z\right\} {= \lim_{s\to\infty} \log q^*_{\tube,s}(\lfloor\alpha s\rfloor)}.
\end{equation}
The function $\log\mathcal L^*_\tube(\alpha)$ is finite and concave for $\alpha\in(2,W)$. 
The inverse Legendre transform is then
\begin{equation}\label{Gs_inverse_Legendre}
\mathcal G^*_\tube(z) = \sup_{2<\alpha<W}\left\{\log\mathcal L^*_\tube(\alpha) + \alpha z\right\}.
\end{equation}

We will determine the behaviour of $\log\mathcal L^*_{\tube}(\alpha)$ as $\alpha\to W^-$, which, together with~\eqref{eqn:sup_z_infty}, informs the behaviour of $\mathcal G^*_\tube(z)$ for $z\to\infty$. For readability we split the result into an upper and lower bound.

\begin{lem}\label{lem:logLs_upperbound}
For any tube size $L\times M$, the density function $\mathcal L^*_\tube(\alpha)$ satisfies
\begin{equation}\label{eqn:logLs_upperbound}
\log \mathcal L^*_\tube(W^-)\equiv \lim_{\alpha\to W^-}\log \mathcal L^*_\tube(\alpha)  \leq \betafull_\tube.
\end{equation}
\end{lem}

\begin{proof}
{The following argument is inspired by a proof of \cite{Rychlewski2011Selfavoiding} regarding adsorbing self-avoiding walks.}

Define
\[j^*_{\tube,s}(m) = \sum_{n\geq m} p^*_{\tube,s}(n),\]
that is, the number of $\mathcal P^*_\tube$ polygons of span $s$ and length at least $m$. 

Given any fixed $r\leq s +1$, we write $s+1=pr+t$ with $0\leq t<r$, and think of a polygon of span $s$ as a connected sequence of $p$ $r$-blocks and (possibly) one $t$-block.
{If a polygon has span $s$ and length $n$ then it has $W(s+1)-n$ unoccupied vertices within its $s+1$ hinges. Letting $u=W(s+1)-m$, the maximum number of unoccupied vertices in a polygon with at least length $m$, and then by considering all possible choices for the number $i$ of $r$-blocks with unoccupied vertices,} we have
\[j^*_{\tube,s}(m) \leq \sum_{i=0}^u \binom{p}{i}\left(b_{\tube,r}\right)^i \left(\bfull_{\tube,r}\right)^{p-i}{b_{\tube,t}}.\]
For any fixed $\delta>0$ take $r$ sufficiently large ($>N_{\delta}$) so that $b_{\tube,r}\leq e^{(\beta_\tube+\delta)r}$ and $\bfull_{\tube,r} \leq e^{(\betafull_\tube+\delta)r}$. Then
\begin{align}
j^*_{\tube,s}(m) &\leq {b_{\tube,t}} \sum_{i=0}^u \binom{p}{i} e^{ir(\beta_\tube+\delta)}e^{(p-i)r(\betafull_\tube+\delta)} \notag\\
&= {b_{\tube,t}}e^{rp(\betafull_\tube+\delta)}\sum_{i=0}^u \binom{p}{i}e^{ir(\beta_\tube-\betafull_\tube)}.\label{eqn:js_upperbound}
\end{align}
Now let $m=\lfloor\alpha s\rfloor$, so that $u=W(s+1) -\lfloor \alpha s\rfloor$. Noting that $p\sim s/r$, take $\alpha$ sufficiently close to $W$ so that $u<p/2$ {($\alpha > W-1/(2r+4)$ is sufficient)}. Then the largest summand of~\eqref{eqn:js_upperbound} is the last one, so
\[j^*_{\tube,s}(\lfloor \alpha s\rfloor) \leq {b_{\tube,t} } e^{rp(\betafull_\tube+\delta)}(u+1)\binom{p}{u} e^{ru(\beta_\tube - \betafull_\tube)}.\]
Take logs, divide by $s$ and apply Stirling's {formula}:
\begin{multline*}
\frac1s \log j^*_{\tube,s}(\lfloor \alpha s \rfloor) \leq \frac1s \log {b_{\tube,t}} + \frac{rp(\betafull_\tube+\delta)}{s} + \frac{ru(\beta_\tube-\betafull_\tube)}{s} + \frac1s \log(u+1) \\
-\frac{p}{s}\log\left(\frac{p-u}{p}\right) + \frac{u}{s}\log\left(\frac{p-u}{u}\right) + O\left(\frac{\log s}{s}\right).
\end{multline*}
With $r>N_{\delta}$ and $\alpha > W-1/(2r+4)$ fixed, take a $\limsup$ as $p\to\infty$ (and hence $s\to\infty$) to find
\begin{align*}
\log\mathcal L^*_\tube(\alpha) &\leq \limsup_{s\to\infty}\frac1s \log j^*_{\tube,s}(\lfloor \alpha s\rfloor) \\
&\leq \betafull_\tube + \delta + r(W-\alpha)(\beta_\tube - \betafull_\tube) - \frac1r \log(1-r(W-\alpha)) + (W-\alpha) \log\left(\frac{1}{r(W-\alpha)}-1\right).
\end{align*}
In the limit $\alpha\to W^-$,
\[\log \mathcal L^*_\tube(W^-) \leq \betafull_\tube + \delta.\]
Since $\delta$ can be arbitrarily small, the proof is complete.
\end{proof}

The proof of the other bound makes use of Lemma~\ref{lem:Tn_lowerbound}. 
\begin{lem}\label{lem:logLs_lowerbound}
\[{\log\mathcal L^*_\tube(W^-)\equiv} \lim_{\alpha\to W^-} \log \mathcal L^*_\tube(\alpha)  \geq \betafull_\tube.\]
\end{lem}

\begin{proof}
By definition, any $s$-block or full $s$-block can be `completed', by adding edges at one or both of its ends, to create a self-avoiding polygon of span $\geq s+1$. In particular, there are constants $s_0$ and $n_0$ (dependant on the dimensions of the tube $\tube$) such that any full $s$-block can be completed into a unique $\mathcal P^*_\tube$ polygon of span $s+s_0$ and length between $Ws$ and $Ws+n_0$. So
\[\bfull_{\tube,s} \leq \sum_{n=Ws}^{Ws+n_0} q^*_{\tube,s+s_0}(n).\]
Now let $n^\mathrm{max}_{s+s_0}$ be the value of $n$ between $Ws$ and $Ws+n_0$ which maximises $q^*_{\tube,s+s_0}(n)$ (if there are multiple such values, take the smallest one). We then have
\[\bfull_{\tube,s}\leq (n_0+1)q^*_{\tube,s+s_0}\left(n^\mathrm{max}_{s+s_0}\right).\]
Observe that $n^\mathrm{max}_s$ is a sequence which satisfies the conditions of Lemma~\ref{lem:Tn_lowerbound}: it is by definition a value between the minimum and maximum lengths for $\mathcal P^*_\tube$ polygons of span $s$, and $n^\mathrm{max}_s = Ws + o(s)$. So
\begin{align*}
\log\mathcal L^*_\tube(W^-) &\geq \limsup_{s\to\infty} \frac1s \log q^*_{\tube,s}\left(n^\mathrm{max}_s\right) \\
&\geq \limsup_{s\to\infty}\frac1s \log \left(\frac{\bfull_{\tube,s-s_0}}{n_0+1}\right) \\
&= \lim_{s\to\infty}\frac1s \log \bfull_{\tube,s} \\
&= \betafull_\tube.\qedhere
\end{align*}
\end{proof}

Now Lemmas~\ref{lem:logLs_upperbound} and~\ref{lem:logLs_lowerbound}, together with~\eqref{Gs_inverse_Legendre} and~\eqref{eqn:sup_z_infty}, imply the following.

\begin{cor}\label{cor:Gs_asymp}
In the limit as $z\to\infty$, the free energy $\mathcal G^*_\tube(z)$ is asymptotic to $Wz+\betafull_\tube$. That is,
\[\lim_{z\to\infty}\left(\mathcal G^*_\tube(z) - Wz\right) = \betafull_\tube.\]
\end{cor}

We are now able to complete the proof of the main theorem of this section.

\begin{proof}[Proof of Theorem~\ref{thm:f_minus_infinity}]
For given rational $\alpha\in(2,W)$, we have
\[\log \mathcal L^*_\tube(\alpha) = \lim_{s\to\infty}\frac1s \log q^*_{\tube,s}(\lfloor \alpha s\rfloor).\]
If we take this limit through values of $s$ such that $s/\alpha$ is an integer, then this can be written as
\begin{align*}
\log \mathcal L^*_\tube(\alpha) &= \lim_{s\to\infty} \frac{1}{s/\alpha} \log q^*_{\tube,s/\alpha}(s) \\
&= \lim_{s\to\infty}\frac{\alpha}{s} \log p^*_{\tube,s}(s/\alpha) \\
&= \alpha \log \mathcal S^*_\tube(1/\alpha).
\end{align*}
Continuity allows us to extend this result to all $\alpha\in(2,W)$, and it can alternatively be written as
\begin{equation}\label{eqn:logL_vs_logS}
\epsilon \log \mathcal L^*_\tube(1/\epsilon) = \log \mathcal S^*_\tube(\epsilon)
\end{equation}
for $\epsilon\in(1/W,1/2)$.

Now consider~\eqref{eqn:Fstube_inverse_Legendre} in the case that $f\to-\infty$. By~\eqref{eqn:sup_z_minus_infty}, the behaviour of $\mathcal F^*_\tube(f)$ in this limit will be determined by the behaviour of $\log\mathcal S^*_\tube(\epsilon)$ as $\epsilon\to (1/W)^+$. 
By~\eqref{eqn:logL_vs_logS} and Lemmas~\ref{lem:logLs_upperbound} and~\ref{lem:logLs_lowerbound},
\begin{align*}
\log \mathcal S^*_\tube\left((1/W)^+\right){\equiv \lim_{\epsilon\to(1/W)^+}   \log \mathcal S^*_\tube(\epsilon) }&= \frac1W \log \mathcal L^*_\tube(W-) \\
&= \frac{\betafull_\tube}{W},
\end{align*}
so that by~\eqref{eqn:sup_z_minus_infty}, $\mathcal F^*_\tube(f)$ is asymptotic to $f/W + \betafull_\tube/W$ as $f\to-\infty$. Since $\mathcal F_\tube(f)=\mathcal F^*_\tube(f)$, the theorem is complete.
\end{proof}

\section{Hamiltonian polygons}\label{sec:hamiltonian}

Theorem~\ref{thm:f_minus_infinity} establishes that, in the limit of a large compressive force, the free energy of polygons in an $L\times M$ tube is related to the growth rate $\betafull_\tube$ of full $s$-blocks in the tube. At first, this may seem peculiar: one might expect that the $f\to-\infty$ asymptote should be related to the growth rate of some easily described class of \emph{polygons}, not \emph{blocks}. In fact we do expect this to be the case. The precise statement of our conjecture, corroborated by numerical analysis for small tube sizes, is presented later in this section (Conjecture~\ref{conj:compressed_ham}).

Recall that, if the first and last half-sections of an $s$-block are empty, the $s$-block itself forms a polygon of span $s-1$. Conversely, any polygon $\pi$ of span $s$ corresponds to a unique $(s+1)$-block. If that $(s+1)$-block is full, we will say that $\pi$ is \emph{Hamiltonian}. Note that, since $\pi$ occupies every vertex in its $s+1$ hinges, it must have length $n=(s+1)W = (s+1)(L+1)(M+1)$. Then because $n$ must be even, we conclude that Hamiltonian polygons of span $s$ can exist only if $W$ is even or $s$ is odd.

Let $\pham_{\tube,n}$ be the number of Hamiltonian polygons of length $n$ in the tube $\tube$, defined up to translation in the $x$-direction. Note that $\pham_{\tube,n}=0$ if $n$ is not a multiple of $W$; moreover, if $W$ is odd then $n$ must be a multiple of $2W$.

The following result establishes that Hamiltonian polygons have a growth rate, and is proved here using arguments adapted from~\cite{Eng2014Thesis}*{Chapter 4}.

\begin{thm}[\cite{Eng2014Thesis}*{Chapter 4}]\label{thm:ham_growthrate}
The limit
\begin{equation}\label{eqn:ham_growthrate}
\kappaham_\tube = \lim_{n\to\infty}\frac1n \log \pham_{\tube,n}
\end{equation}
exists, where the limit is taken through values of $n$ which are multiples of $W$ (resp.~$2W$) when $W$ is even (resp.~odd). The limit is finite.
\end{thm}

The proof of Theorem~\ref{thm:ham_growthrate} will follow from a concatenation argument. Before we begin, it will be convenient to introduce two special hinges, constructed via a process called \emph{zig-zagging}. This process, operating in an $L\times M$ rectangle of the $y$-$z$ plane (i.e.~a hinge of $\tube$, with $0\leq y\leq L$ and $0\leq z\leq M$), generates a self-avoiding walk via the following algorithm.
\begin{enumerate}
\item Begin at initial vertex $(x,y_0,z_0)$.
\item If possible (without violating self-avoidance), take steps in the positive $z$-direction, without passing $z=M$. Go to step 3.
\item If possible (without violating self-avoidance), take steps in the negative $z$-direction, without passing $z=0$. Go to step 4.
\item If possible (without violating self-avoidance or passing $y=L$), take a step in the positive $y$-direction, and return to step 2. If not, terminate the process.
\end{enumerate}

The two special hinges are then defined as follows.
\begin{itemize}
\item $H^A_\tube(x)$ consists of the edges $(x,0,0)\edge(x,1,0)\edge\ldots\edge(x,L,0)$, together with a zig-zagging starting at $(x,0,1)$.
\item $H^B_\tube(x)$ consists of the edges $(x,1,0)\edge(x,2,0)\edge\ldots\edge(x,L,0)$, together with a zig-zagging starting at $(x,0,0)$.
\end{itemize}
See Figure~\ref{fig:hinges_HA_HB} for examples. Note that if $M=0$ then $H^A_\tube(x)$ is just a line of edges from $(x,0,0)$ to $(x,L,0)$, while $H^B_\tube(x)$ is the vertex $(x,0,0)$ together with edges from $(x,1,0)$ to $(x,L,0)$.

\begin{figure}
\centering
\begin{subfigure}[b]{0.4\textwidth}
\centering
\begin{tikzpicture}
\tikzset{scale=0.7}
\draw[step=1cm, gray, thin] (0,0) grid (7,4);
\draw[line width=2pt, line cap=rect] (0,0) -- (7,0);
\draw[line width=2pt, line cap=rect] (0,1) -- (0,4) -- (1,4) -- (1,1) -- (2,1) -- (2,4) -- (3,4) -- (3,1) -- (4,1) -- (4,4) -- (5,4) -- (5,1) -- (6,1) -- (6,4) -- (7,4) -- (7,1);
\end{tikzpicture}
\caption{The hinge $H^A_\tube(x)$ in a $7\times 4$ tube.}
\end{subfigure}
\begin{subfigure}[b]{0.4\textwidth}
\centering
\begin{tikzpicture}
\tikzset{scale=0.7}
\draw[step=1cm, gray, thin] (0,0) grid (6,4);
\draw[line width=2pt, line cap=rect] (0,0) -- (6,0);
\draw[line width=2pt, line cap=rect] (0,1) -- (0,4) -- (1,4) -- (1,1) -- (2,1) -- (2,4) -- (3,4) -- (3,1) -- (4,1) -- (4,4) -- (5,4) -- (5,1) -- (6,1) -- (6,4); 
\end{tikzpicture}
\caption{The hinge $H^A_\tube(x)$ in a $6\times 4$ tube.}
\end{subfigure}

\vskip0.5cm

\begin{subfigure}[b]{0.4\textwidth}
\centering
\begin{tikzpicture}
\tikzset{scale=0.7}
\draw[step=1cm, gray, thin] (0,0) grid (7,4);
\draw[line width=2pt, line cap=rect] (1,0) -- (7,0);
\draw[line width=2pt, line cap=rect] (0,0) -- (0,4) -- (1,4) -- (1,1) -- (2,1) -- (2,4) -- (3,4) -- (3,1) -- (4,1) -- (4,4) -- (5,4) -- (5,1) -- (6,1) -- (6,4) -- (7,4) -- (7,1);
\end{tikzpicture}
\caption{The hinge $H^B_\tube(x)$ in a $7\times 4$ tube.}
\end{subfigure}
\begin{subfigure}[b]{0.4\textwidth}
\centering
\begin{tikzpicture}
\tikzset{scale=0.7}
\draw[step=1cm, gray, thin] (0,0) grid (6,4);
\draw[line width=2pt, line cap=rect] (1,0) -- (6,0);
\draw[line width=2pt, line cap=rect] (0,0) -- (0,4) -- (1,4) -- (1,1) -- (2,1) -- (2,4) -- (3,4) -- (3,1) -- (4,1) -- (4,4) -- (5,4) -- (5,1) -- (6,1) -- (6,4); 
\end{tikzpicture}
\caption{The hinge $H^B_\tube(x)$ in a $6\times 4$ tube.}
\end{subfigure}
\caption{Hinges $H^A_\tube(x)$ and $H^B_\tube(x)$ in the $y$-$z$ plane, when $L$ is odd or even. The bottom left corner in each is the vertex $(x,y,z) = (x,0,0)$.}
\label{fig:hinges_HA_HB}
\end{figure}
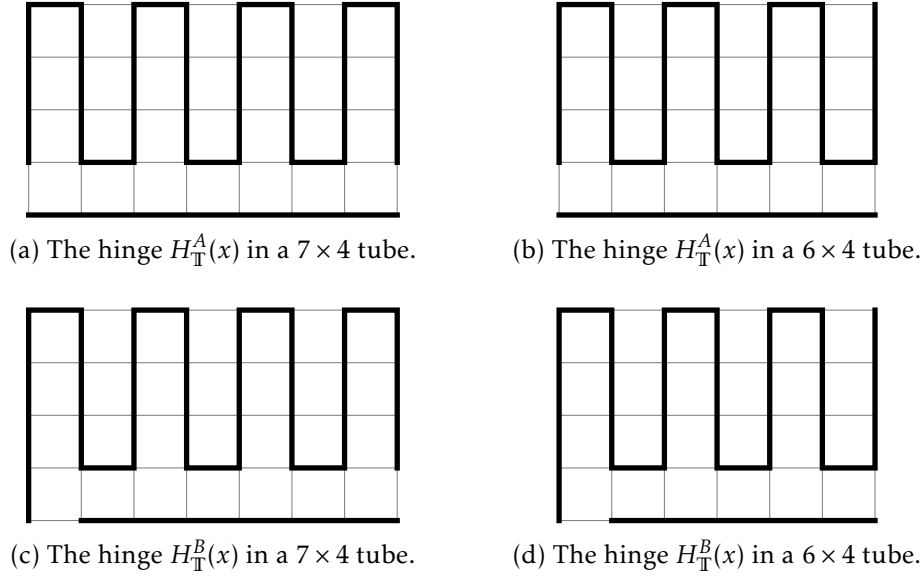

\begin{proof}[Proof of Theorem~\ref{thm:ham_growthrate}]
We will show that $\pham_{\tube,n}$ is a supermultiplicative sequence, by demonstrating that any two Hamiltonian polygons in $\tube$ can be concatenated to give a third. 

Let $\pi$ be a Hamiltonian polygon in $\tube$ of length $n$ and span $s$. Since $\pi$ is Hamiltonian, the vertex $(0,0,0)$ must be occupied, and thus at least one of the edges $(0,0,0)\edge(0,1,0)$ and $(0,0,0)\edge(0,0,1)$ must also be occupied. (Clearly if $M=0$ then it must be the former.) We say that $\pi$ is of \emph{type} $S_1$ if $(0,0,0)\edge(0,1,0)$ is occupied, otherwise it is of \emph{type} $S_0$. Similarly, at least one of the edges $(s,0,0)\edge(s,1,0)$ and $(s,0,0)\edge(s,0,1)$ must be occupied by $\pi$; if the former is occupied then $\pi$ is of \emph{type} $F_1$, otherwise it is of \emph{type} $F_0$. (The $S$ and $F$ stand for \emph{start} and \emph{finish}.)

Now let $\pi_1$ and $\pi_2$ be two Hamiltonian polygons in $\tube$, of lengths $n_1$ and $n_2$ and spans $s_1$ and $s_2$ respectively. We will define a new polygon $\pi_1\circ\pi_2$ generated by concatenation. There are four cases to consider, depending on whether $\pi_1$ is of type $F_0$ or $F_1$, and whether $\pi_2$ is of type $S_0$ or $S_1$. In all cases, we begin by translating $\pi_2$ a distance of $s_1+3$ in the positive $x$-direction.
\begin{itemize}
\item[(a)] $(\pi_1,\pi_2)$ of types $(F_1,S_1)$: Insert hinges $H^B_\tube(s_1+1)$ and $H^B_\tube(s_1+2)$. Delete edges $(s_1,0,0)\edge(s_1,1,0)$ in $\pi_1$ and $(s_1+3,0,0)\edge(s_1+3,1,0)$ in (the translation of) $\pi_2$. Insert the two edges required to join $\pi_1$ to $H^B_\tube(s_1+1)$, the two edges required to join $H^B_\tube(s_1+2)$ to $\pi_2$, and the two edges required to join $H^B_\tube(s_1+1)$ to $H^B_\tube(s_1+2)$.
\item[(b)] $(\pi_1,\pi_2)$ of types $(F_1,S_0)$: Insert hinges $H^B_\tube(s_1+1)$ and $H^A_\tube(s_1+2)$. Delete edges $(s_1,0,0)\edge(s_1,1,0)$ in $\pi_1$ and $(s_1+3,0,0)\edge(s_1+3,0,1)$ in $\pi_2$. Insert the three pairs of edges required to join $\pi_1$ to $H^B_\tube(s_1+1)$, $H^A_\tube(s_1+2)$ to $\pi_2$, and $H^B_\tube(s_1+1)$ to $H^A_\tube(s_1+2)$.
\item[(c)] $(\pi_1,\pi_2)$ of types $(F_0,S_1)$: Insert hinges $H^A_\tube(s_1+1)$ and $H^B_\tube(s_1+2)$. Delete edges $(s_1,0,0)\edge(s_1,0,1)$ in $\pi_1$ and $(s_1+3,0,0)\edge(s_1+3,1,0)$ in $\pi_2$. Insert the three pairs of edges required to join $\pi_1$ to $H^A_\tube(s_1+1)$, $H^B_\tube(s_1+2)$ to $\pi_2$, and $H^A_\tube(s_1+1)$ to $H^B_\tube(s_1+2)$.
\item[(d)] $(\pi_1,\pi_2)$ of types $(F_0,S_0)$: Insert hinges $H^A_\tube(s_1+1)$ and $H^A_\tube(s_1+2)$. Delete edges $(s_1,0,0)\edge(s_1,0,1)$ in $\pi_1$ and $(s_1+3,0,0)\edge(s_1+3,0,1)$ in $\pi_2$. Insert the three pairs of edges required to join $\pi_1$ to $H^A_\tube(s_1+1)$, $H^A_\tube(s_1+2)$ to $\pi_2$, and $H^A_\tube(s_1+1)$ to $H^A_\tube(s_1+2)$.
\end{itemize}

\begin{figure}[t]
\centering
\begin{subfigure}[b]{\textwidth}
\centering
%\begin{tikzpicture}
%\tikzset{scale=0.5, every node/.style={circle, fill, inner sep=2.5pt}}
%\begin{knot}[clip width=3]
%\strand[ultra thick] (0,0) node{} -- (0,3) node{} -- (1,4) node{} -- (1,1) node{} -- (2,2) node{} -- (2,5) node{} -- (5,5) node{} -- (8,5) node{} -- (7,4) node{} -- (7,1) node{} -- (8,2) node{};
%\strand[ultra thick] (8,2) -- (5,2) node{} -- (4,1) node{} -- (4,4) node{} -- (3,3) node{};
%\strand[ultra thick] (3,3) -- (6,3) node{} -- (6,0) node{} -- (3,0) node{} -- (0,0);
%\strand[ultra thick] (12,0) node{} -- (13,1) node{} -- (13,4) node{} -- (14,5) node{} -- (14,2) node{} -- (17,2) node{} -- (20,2) node{} -- (19,1) node{};
%\strand[ultra thick] (19,1) -- (19,4) node{} -- (20,5) node{} -- (17,5) node{} -- (16,4) node{} -- (16,1) node{} -- (15,0) node{} -- (18,0) node{} -- (18,3) node{};
%\strand[ultra thick] (18,3) -- (15,3) node{} -- (12,3) node{} -- (12,0);
%\flipcrossings{2,3,4,5,6,7,8,9}
%\end{knot}
%\tikzset{every node/.style={}}
%\draw (10,2) node{\LARGE $\circ$};
%\end{tikzpicture}
\includegraphics{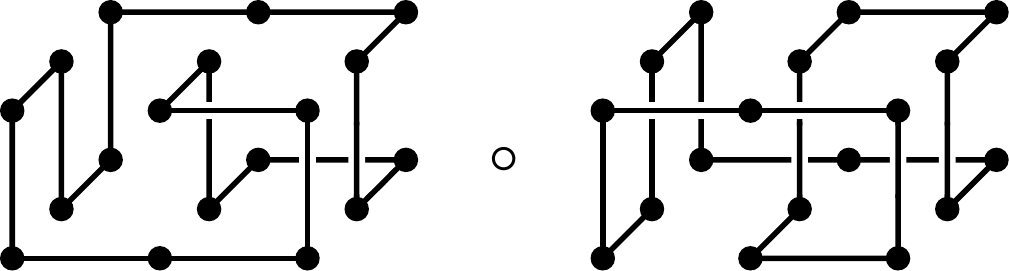}
\end{subfigure}

\vskip1cm

\begin{subfigure}[b]{\textwidth}
\centering
%\begin{tikzpicture}
%\tikzset{scale=0.5}
%\draw (0,2) node{\LARGE $=$};
%\tikzset{every node/.style={circle, fill, inner sep=2.5pt}}
%\foreach \x in {2,5,...,23}
%	\foreach \y in {0,1,2}
%		\foreach \z in {0,3}
%			\draw (\x+\y, \y+\z) node{};
%\begin{scope}[on background layer]
%\begin{knot}[clip width=3]
%\strand[ultra thick] (2,0) -- (2,3) -- (3,4) -- (3,1) -- (4,2) -- (4,5) -- (7,5) -- (10,5) -- (9,4) -- (9,1) -- (10,2);
%\strand[ultra thick] (10,2) -- (7,2) -- (6,1) -- (6,4) -- (5,3);
%\strand[ultra thick] (5,3) -- (8,3);
%\strand[ultra thick] (8,0) -- (5,0) -- (2,0);
%\strand[ultra thick] (18,1) -- (18,4) -- (19,5) -- (19,2) -- (22,2) -- (25,2) -- (24,1);
%\strand[ultra thick] (24,1) -- (24,4) -- (25,5) -- (22,5) -- (21,4) -- (21,1) -- (20,0) -- (23,0) -- (23,3);
%\strand[ultra thick] (23,3) -- (20,3) -- (17,3) -- (17,0);
%\strand[ultra thick, red] (11,0) -- (12,1) -- (13,2);
%\strand[ultra thick, red] (11,3) -- (12,4) -- (13,5);
%\strand[ultra thick, red] (15,1) -- (16,2);
%\strand[ultra thick, red] (14,0) -- (14,3) -- (15,4) -- (16,5);
%\strand[ultra thick, blue] (8,0) -- (11,0);
%\strand[ultra thick, blue] (8,3) -- (11,3);
%\strand[ultra thick, blue] (13,2) -- (16,2);
%\strand[ultra thick, blue] (13,5) -- (16,5);
%\strand[ultra thick, blue] (14,0) -- (17,0);
%\strand[ultra thick, blue] (15,1) -- (18,1);
%\flipcrossings{2,3,4,5,6,7,8,9}
%\end{knot}
%\end{scope}
%\end{tikzpicture}
\includegraphics{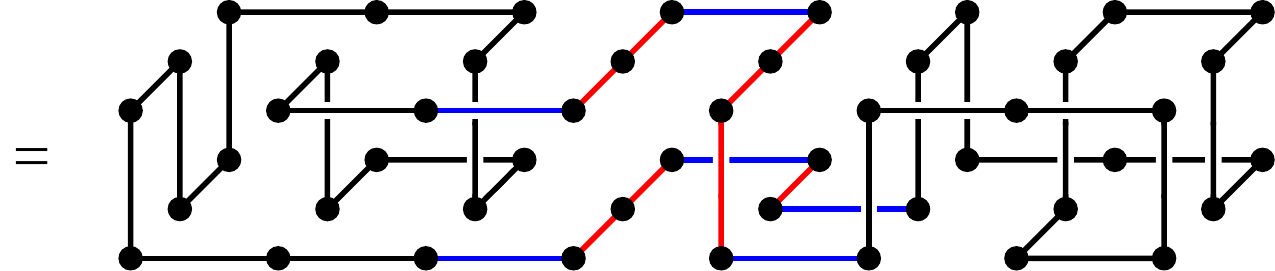}
\end{subfigure}
\caption{The concatenation operation of two Hamiltonian polygons in the $2\times1$ tube, as described in the proof of Theorem~\ref{thm:ham_growthrate}. This is case (c), where the first polygon is of type $F_0$ and the second is of type $S_1$. The red edges are the two special hinges $H^A_\tube(3)$ and $H^B_\tube(4)$, and the blue edges connect these special hinges to the two polygons and to each other.}
\label{fig:ham_concatenation1}
\end{figure}

See Figure~\ref{fig:ham_concatenation1} for an example. In each of these four cases, we have constructed a unique Hamiltonian polygon of length $n_1+n_2+2W$ and span $s_1+s_2+3$. We thus have
\begin{equation}\label{eqn:ham_submult}
\pham_{\tube,n_1}\pham_{\tube,n_2} \leq \pham_{\tube,n_1+n_2+2W}.
\end{equation}
Subtracting $2W$ from each of $n_1$ and $n_2$ gives
\[\pham_{\tube,n_1-2W}\pham_{\tube,n_2-2W} \leq \pham_{\tube,n_1+n_2-2W},\]
so that $\left\{\log \pham_{\tube,n-2W}\right\}$ is a subadditive sequence. It follows that the limit~\eqref{eqn:ham_growthrate} exists. Moreover, it is straightforward to connect up sequences of $H^A_\tube(x)$ hinges (or alternatively, sequences of $H^B_\tube(x)$ hinges) in order to show that, for any $n$ a multiple of $W$ (resp.~$2W$) when $W$ is even (resp.~$W$ is odd), there exists a Hamiltonian polygon of length $n$. So for those values of $n$,
\[1 \leq \pham_{\tube,n} \leq p_{\tube,n} \quad \implies \quad 0 \leq \kappaham_\tube \leq \mathcal F_\tube(0)<\infty.\qedhere\]
\end{proof}

As with general polygons in $\tube$, one can associate a force $f$ with the span of Hamiltonian polygons, to obtain a partition function $\Zham_{\tube,n}(f)$. Moreover, since all Hamiltonian polygons of length $n$ have the same span $s=n/W-1$, we have
\[\Zham_{\tube,n}(f) = \pham_{\tube,n}e^{f(n/W-1)}.\]
The corresponding free energy then has a simple form:
\[\Fham_{\tube}(f) = \lim_{n\to\infty} \frac1n \log \Zham_{\tube,n}(f) = \kappaham_\tube + \frac{f}{W},\]
where the limit is taken through values of $n$ which are multiples of $W$ or $2W$ as appropriate.

Having established the existence of a growth rate $\kappaham_\tube$ and free energy $\Fham_\tube(f)$, we are now able to state the conjectured relationship between compressed and Hamiltonian polygons.

\begin{conj}\label{conj:compressed_ham}
Hamiltonian polygons and full $s$-blocks in the $L\times W$ tube $\tube$, counted by length instead of span, have the same growth rate. That is,
\[\kappaham_\tube = \frac{\betafull_\tube}{W}\]
where $W=(L+1)(M+1)$. Consequently, in the limit $f\to-\infty$, the free energy $\mathcal F_\tube(f)$ of polygons in the tube is asymptotic to $\Fham_\tube(f) = \kappaham_\tube+f/W$. That is,
\[\lim_{f\to-\infty} \left(\mathcal F_\tube(f) - \Fham_\tube(f)\right) = 0.\]
\end{conj}

We next explore the validity of this conjecture for small tube sizes using transfer matrix calculations.

\subsection{Transfer-matrices and Hamiltonian polygons}\label{subsec:hamiltonian}

We focus first on defining  $1$-patterns in terms of $1$-blocks, and then use 1-patterns to define a transfer matrix. 
To do this, first consider any $\omega\in {\cal P}_{\tube}$ and let $s$ be its span.  The polygon $\omega$ uniquely defines a sequence of $s+1$ connected 1-blocks: $E_0(\omega), E_1(\omega), ...,E_s(\omega)$.  
Given a
 $j\in\{1,...,s\}$,  $\omega$ can be thought of as a connected sequence of three embeddings $E_j^1$, $E_j(\omega)$ and $E_j^3$ where $E_j^1$ (resp.~$E_j^3$) consists of the edges and half-edges of $\omega$ before (resp.~after) the plane $x=j-\frac{1}{2}$ ($x=j+\frac{1}{2}$). 
Since $\omega$ is a polygon, the vertices  of $E_j(\omega)$ in the plane $x=j-\frac{1}{2}$ are connected pairwise by sequences of edges in $E_j^1$.
To define a  $1$-pattern, it is unnecessary to keep the full details of these edge sequences; rather, it will be enough to store the connectivity information in terms of which of the left-most vertices of  $E_j(\omega)$
are connected together in $E_j^1$.  
For this, we first label the vertices of the left-most plane of $E_j(\omega)$ lexicographically as $v^j_1,...,v^j_{r_j}$. Next we obtain a pair-partition ${\cal S}_j$  of the vertex labels $\{1,...,r_j\}$ from $V^{j}=\{v^j_1,...,v^j_{r_j}\}$, using the connectivity information from $E_j^1$.  
We then define the {\it left connectivity information} for  $E_j(\omega)$ by this pair partition ${\cal E}_j={\cal S}_j$.  
For $E_0(\omega)$,  because its left-connectivity information is completely determined by the 1-block we define its left-connectivity information to be ${\cal E}_{0}=\phi$, the empty set.
Now $\omega$'s $j$th \emph{proper $1$-pattern} is defined to be the ordered pair
$\omega_j=({\cal E}_j,E_j(\omega))$, $j=1,...,s-1$; its \emph{right-most $1$-pattern}, the ordered pair $\omega_s=({\cal E}_s,E_s(\omega))$; and its \emph{left-most $1$-pattern},  $\omega_0=({\cal E}_0,E_0(\omega))$.
Hence $\omega$ generates a unique sequence of $1$-patterns $(\omega_0,\omega_1,...,\omega_{s-1},\omega_s)$ and, for convenience, we write $\omega=(\omega_0,\omega_1,...,\omega_{s-1},\omega_s)$.
From this we can define ${\cal A}_1$, ${\cal A}_2$, and ${\cal A}_3$, respectively, as the set of all distinct (up-to $x$-translation)  left-most, proper,  and right-most  $1$-patterns that result from some $\omega\in{\cal P}_{\tube}$ with span $s\geq 1$.  We also define ${\cal A}_0$
to be the set of all $\omega\in{\cal P}_{\tube}$ with span $s=0$.

Given two $1$-patterns $\pi_1=({S}_{1,1},E_1)$, and $\pi_2=({S}_{2,1},E_2)$,  we consider whether $E_1$ followed by $E_2$ is a possible  $2$-block of a  polygon.   Note that  $S_{1,1}$ and $E_1$ induce a pair partitioning for the vertices in the right-most plane of $E_1$, call this pair partition ${ S}_{1,2}$.  We thus
say that $\pi_2$ can follow $\pi_1$ (or equivalently, $\pi_1$ can precede $\pi_2$) if ${ S}_{1,2}={ S}_{2,1}$ and the right-most plane of  $E_1$ is the same as the left-most plane of $E_2$.  (Note that  we are allowing $\pi_1$ to be a left-most pattern or $\pi_2$ to be a right-most pattern.)
We say a sequence of $1$-patterns, $\pi_1,\pi_2, ...., \pi_r$,  is {\it properly connected}  if $\pi_{i+1}$ can follow $\pi_i$ for each $i=1,...,r-1$.  We refer to the entire sequence $\pi_1,\ldots,\pi_r$ as an \emph{$r$-pattern}. Let $t_{\tube,r}$ be the number of $r$-patterns in the tube $\tube$, and let $t^\mathrm{F}_{\tube,r}$ be the number of $r$-patterns whose underlying $r$-blocks are full. We refer to the latter as \emph{full $r$-patterns}.  Any $r$-pattern which consists of a sequence of proper $1$-patterns is called a \emph{proper $r$-pattern}.
{By definition, for each $\omega\in{\cal P}_{\tube}$ (or ${\cal P}_{\tube}^\mathrm{H}$, the subset of Hamiltonian polygons)
its sequence $\omega_0,\omega_1,....,\omega_{s-1},\omega_s$ of 1-patterns gives an $(s+1)$-pattern (a full $(s+1)$-pattern),  and  for any $r\geq 2$, each $r$-pattern (or full $r$-pattern) starting with a left-most 1-pattern (full left-most 1-pattern) and ending with a  right-most 1-pattern (full right-most 1-pattern) yields an
element of ${\cal P}_{\tube}$ (${\cal P}_{\tube}^\mathrm{H}$) . }

\begin{lem}\label{lem:patterns_vs_blocks}
Both $r$-patterns and full $r$-patterns have exponential growth rates, and these are equal to $\beta_\tube$ and $\beta^\mathrm{F}_\tube$ respectively.
\end{lem}

\begin{proof}
Patterns are distinguished from blocks by the inclusion of left connectivity information. Each $r$-pattern corresponds to a unique $r$-block, but an $r$-block $\omega$ may correspond to multiple $r$-patterns, as there may be multiple valid sets of left connectivity information which can be matched to $\omega$. However, observe that the number of valid sets of left connectivity information is bounded above by a function of the tube size; namely, the number of pair partitions of $W$ (if $W$ is even) or $W-1$ (if $W$ is odd) vertices. This number is $(W-1)!!$ if $W$ is even and $(W-2)!!$ if $W$ is odd. Hence
\[b_{r,\tube} \leq t_{r,\tube} \leq (W-1)!! \,b_{r,\tube}.\]
Take logs, divide by $r$ and take $r\to\infty$, to find
\[\lim_{r\to\infty} \frac1r \log t_{\tube,r} = \beta_\tube.\]
Exactly the same arguments apply to full $r$-patterns, and we have
\[\lim_{r\to\infty} \frac1r \log t^\mathrm{F}_{\tube,r} = \beta^\mathrm{F}_\tube.\qedhere\]
\end{proof}

With this definition of patterns, we can follow the approaches used in \cite{Eng2014Thesis}  to obtain
transfer matrices.    We will focus on full patterns, and hence define four sets  ${\cal A}_0^\mathrm{F}$, ${\cal A}_1^\mathrm{F}$, ${\cal A}_2^\mathrm{F}$, and ${\cal A}_3^\mathrm{F}$ corresponding, respectively, to those elements of ${\cal A}_0$, ${\cal A}_1$, ${\cal A}_2$, and ${\cal A}_3$ which are full.  We assign a labelling to the elements of $\cup_{k=0}^3{\cal A}_k^\mathrm{F}$ and denote them as $\pi_1,\pi_2, ...., \pi_{r_{\tube}}$.   Then we obtain the $r_{\tube}\times r_{\tube}$ transfer matrix $T^\mathrm{F}(x)$ for full $1$-patterns as follows:
\begin{equation*}
\left[T^\mathrm{F}(x)\right]_{i,j} = \begin{cases} x^{n_{\pi_i} + n_{\pi_j}}=x^W & \text{if $\pi_j$ can follow $\pi_i$} \\ 0 & \text{otherwise,}\end{cases}
\end{equation*}
where $n_{\pi}$ is the length of the 1-block from which the 1-pattern $\pi$ was derived, which is $W$ for full 1-blocks.

The generating function for full patterns can be expressed in terms of this transfer matrix as follows:
\begin{align*}
G^\mathrm{F}(x)=\sum_{s\geq 1} t^\mathrm{F}_{\tube,s} x^{sW} &=  t^\mathrm{F}_{\tube,1}x^W+x^W\sum_{i,j}\left[\sum_{t\geq 0} T^\mathrm{F}(x) \left(T^\mathrm{F}(x)\right)^t\right]_{i,j}\\
&=t^\mathrm{F}_{\tube,1}x^W + x^W\sum_{i,j}\left[  T^\mathrm{F}(x)(I-T^\mathrm{F}(x))^{-1}\right]_{i,j},
\end{align*}
where $ t^\mathrm{F}_{\tube,1}=r_{\tube}$.
The radius of convergence of $G^\mathrm{F}(x)$ is given by $e^{-\beta_{\tube}^\mathrm{F}/W}$  and can also be determined by the smallest value of $x>0$ which satisfies $\det(I-T^\mathrm{F}(x))=\det(I-x^WT^\mathrm{F}(1))=0$ or equivalently $\det(x^{-W}I-T^\mathrm{F}(1))=0$, that is, it is given by the largest eigenvalue of $T^\mathrm{F}(1)$.    
The generating function for Hamiltonian polygons can also be expressed in terms of this transfer matrix as follows:
\begin{align*}
G^\mathrm{H}(x)=\sum_{s\geq 0} p^\mathrm{H}_{\tube,s} x^{(s+1)W} &= \left|{\cal A}_0^\mathrm{F}\right|x^W+p^\mathrm{H}_{\tube,1}x^{2W}+x^W\sum_{i,j}\left[\sum_{t\geq 0} A^\mathrm{H}(x) \left(T^\mathrm{F}(x)\right)^tB^\mathrm{H}(x)\right]_{i,j} \\
&=\left|{\cal A}_0^\mathrm{F}\right|x^W + p^\mathrm{H}_{\tube,1}x^{2W}+x^W\sum_{i,j}\left[  A^\mathrm{H}(x)(I-T^\mathrm{F}(x))^{-1}B^\mathrm{H}(x)\right]_{i,j} ,
\end{align*}
where the matrices $A^\mathrm{H}(x)$ (resp.~$B^\mathrm{H}(x)$) are obtained by first labelling the elements of ${\cal A}^\mathrm{F}_1$ (${\cal A}^\mathrm{F}_3$) as $\pi_{1,1},\pi_{1,2}, ...., \pi_{1,r_{1,\tube}}$ ($\pi_{3,1},\pi_{3,2},\ldots, \pi_{3,r_{3,\tube}}$)
and then, for each $j=1,...,r_{\tube}$: the  $i,j$ entry of $A^\mathrm{H}(x)$ is $x^{W}$ if  $\pi_j$ can follow $\pi_{1,i}$ (0 otherwise), $i=1,...,r_{1,\tube}$; and the
$j,i$ entry of $B^\mathrm{H}(x)$ is $x^{W}$ if  $\pi_{3,i}$ can follow $\pi_j$ (0 otherwise), $i=1,...,r_{3,\tube}$.  
We explain next that  determining whether or not
the conjecture holds is equivalent to determining whether or not the largest eigenvalue of $T^\mathrm{F}(1)$ gives the radius of convergence for $G^\mathrm{H}(x)$.

For two 1-patterns $\pi_i$ and $\pi_j$ in ${\cal A}^\mathrm{F}=\cup_{k=0}^3{\cal A}_k^\mathrm{F}$, we say $\pi_j$ is reachable from $\pi_i$  if  {for some $r$ there is a full $r$-pattern} that starts with $\pi_i$ and ends with $\pi_j$.  $T^\mathrm{F}(x)$ is the weighted adjacency matrix for a directed graph $D^\mathrm{F}$ on the set of elements of ${\cal A}^\mathrm{F}$, and if $\pi_j$ is reachable from $\pi_i$ then there is a directed path from $\pi_i$ to $\pi_j$ in $D^\mathrm{F}$.  We say $\pi_i$ and $\pi_j$ \emph{communicate}  if $\pi_j$ is reachable from $\pi_i$ and $\pi_i$ is reachable from $\pi_j$.  Communication is an equivalence relation which partitions ${\cal A}^\mathrm{F}$ into communication classes that correspond to the strongly connected components of the digraph $D^\mathrm{F}$.  The elements of ${\cal A}^\mathrm{F}$ can then be relabelled in such a way that $T^\mathrm{F}(x)$ is a block upper triangular matrix where the  block matrices along the diagonal are the weighted adjacency matrices for the strongly connected components of $D^\mathrm{F}$ {(this gives the Frobenius normal form of $T^\mathrm{F}(x)$)}.  Hence the
characteristic polynomial of $T^\mathrm{F}(x)$ is the product of the characteristic polynomials of the weighted adjacency matrices for the strongly connected components of $D^\mathrm{F}$. (See for example~\cite[p29-7 and p27-6]{Hogben2013} or \cite[Chapter 3]{Brualdi_1991}.)

We define the Hamiltonian 1-patterns to be those elements of ${\cal A}_2^\mathrm{F}$ which can be part of a Hamiltonian polygon; call this subset   ${\cal A}_2^\mathrm{H}$.  Note that by definition every element of ${\cal A}_2^\mathrm{H}$ is reachable from some element of ${\cal A}_1^\mathrm{F}$.  Further, if we consider any two elements $\pi_i$ and $\pi_j$ in ${\cal A}_2^\mathrm{H}$, then there exists a Hamiltonian polygon $\omega_1$ which contains $\pi_i$ and another Hamiltonian polygon $\omega_2$ which contains $\pi_j$.  The concatenation construction defined earlier in this section can be used to concatenate polygon $\omega_1$ to $\omega_2$ (or vice versa) to create a new Hamiltonian polygon with $\pi_j$ reachable from $\pi_i$ ($\pi_i$ reachable from $\pi_j$) through elements of ${\cal A}_2^\mathrm{H}$.  Thus the subdigraph of $D^\mathrm{F}$ generated by the elements of ${\cal A}_2^\mathrm{H}$ forms a strongly connected digraph $D^\mathrm{H}$ in which every 1-pattern is reachable from every other.  We claim further that this subdigraph is a strongly connected component of $D^\mathrm{F}$ (i.e. it is a maximal strongly connected digraph).  Suppose to the contrary that there exists a larger strongly connected subdigraph of $D^\mathrm{F}$, call it $D$, which contains $D^\mathrm{H}$ as a proper subdigraph.  Let  $\pi_i$ be in the vertex set of $D$ but not in ${\cal A}_2^\mathrm{H}$,  then $\pi_i$ does not occur in a Hamiltonian polygon, however, $\pi_i$ communicates with every vertex of $D^\mathrm{H}$.  A contradiction results by taking a Hamiltonian polygon $\omega$ which contains $\pi_j\in {\cal A}_2^\mathrm{H}$ and inserting at  $\pi_j$ a sequence of properly connected 1-patterns from $\pi_j$ to $\pi_i$ and then from $\pi_i$ to $\pi_j$ to create a Hamiltonian polygon that contains $\pi_i$.  Thus $D^\mathrm{H}$ is a strongly connected component of $D^\mathrm{F}$.  We call its weighted adjacency matrix the Hamiltonian 1-pattern transfer matrix $T^\mathrm{H}(x)$ and it is obtained by restricting $T^\mathrm{F}(x)$ (all other rows and columns removed) to the elements of ${\cal A}_2^\mathrm{H}$.   Thus we also have:
\begin{align*}
G^\mathrm{H}(x)=\sum_{s\geq 0} p^\mathrm{H}_{\tube,s} x^{(s+1)W} &= \left|{\cal A}_0^\mathrm{F}\right|x^W+p^\mathrm{H}_{\tube,1}x^{2W}+x^W{\sum_{i,j}\left[\sum_{t\geq 0} A^\mathrm{H*}(x) \left(T^\mathrm{H}(x)\right)^tB^\mathrm{H*}(x)\right]_{i,j}} \\
&=\left|{\cal A}_0^\mathrm{F}\right|x^W + p^\mathrm{H}_{\tube,1}x^{2W}+ x^W\sum_{i,j}\left[  A^\mathrm{H*}(x)(I-T^\mathrm{H}(x))^{-1}B^\mathrm{H*}(x)\right]_{i,j} ,
\end{align*}
where $A^\mathrm{H*}(x)$ and  $B^\mathrm{H*}(x)$ are obtained from $A^\mathrm{H}(x)$ and  $B^\mathrm{H}(x)$, respectively, by restricting to ${\cal A}_2^\mathrm{H}$, and $T^\mathrm{H}(x)$ will be one of the block matrices along the diagonal in the Frobenius normal form of $T^\mathrm{F}(x)$.
Thus $\det(I-T^\mathrm{F}(x))=\det(I-T^\mathrm{H}(x))\prod_{k\geq 1} \det(I-T_k(x))$ where {$T_k(x), k\geq 1$} are the weighted adjacency matrices for the other strongly connected components of $D^\mathrm{F}$.  The component which corresponds to the smallest root will yield the radius of convergence of $G^\mathrm{F}(x)$.
The conjecture is that this root comes from $\det(I-T^\mathrm{H}(x))=\det(I-x^WT^\mathrm{H}(1))$  and this corresponds to $T^\mathrm{H}(1)$ having the largest eigenvalue $\det(x^{-W}I-T^\mathrm{H}(1))$.  For small tube sizes we have verified this conjecture by determining the strongly connected components and their corresponding adjacency matrices and determining which component(s) determine the radius of convergence.   Table \ref{evidencetable} shows the results.
{In addition to the numerical verifications provided in Table \ref{evidencetable}, in two dimensions (that is, when $M=0$) this conjecture has been verified exactly for $L\leq 5$.}

\begin{table}[phtb]
\newcolumntype{C}{>{\centering\arraybackslash} m{0.13\textwidth} }
\centering
\begin{tabular}{| C | C | C || C | C | C |}
\hline
$ \tube ~\mbox{size} ~L\times M$ & $\kappa_{\tube}^\mathrm{H} $ & next largest growth rate & $ \tube ~\mbox{size} ~L\times M$ & $\kappa_{\tube}^\mathrm{H} $ & next largest growth rate  \\ \hline
$3\times 0$ & 0.232905 & 0 & $1\times 1$ & 0.329239	&	0.173287 \\ \hline
$4\times0$ & 0.239939 & 0.138629 & $2\times 1$  & 0.440750	& 	0.360063 \\ \hline
$5\times0$ & 0.288670 & 0.196889 & $3 \times 1$ & 0.488108	&	0.443274\\ \hline
$6\times0$ & 0.288344 & 0.222048 & $4\times 1$ & 0.515163	&	0.485601 \\ \hline
$7\times0$ & 0.314534 & 0.263113 & $2\times 2$ & 0.516565	&	0.406593 \\ \hline
$8\times0$ & 0.313302 & 0.273317 & & & \\ \hline
\end{tabular}
\caption{Evidence that $\kappa_{\tube}^\mathrm{H} =\beta_{\tube}^\mathrm{F}/W$ for small tube sizes.}
\label{evidencetable}
\end{table}

\section{Summary and Discussion}\label{sec:conclusion}

We have studied a model of self-avoiding polygons restricted to a $L\times M$ rectangular tube $\tube$ of the cubic lattice $\mathbb Z^3$, subject to a force $f$ which acts in a direction parallel to the axis of the tube. {Without loss of generality, we assume $L\geq M\geq 0$ and $L>0$.} When $f>0$ the force effectively stretches the polygons, while when $f<0$ the force is compressive. For all values of $f$ one can define a free energy $\mathcal F_\tube(f)$. We have shown that in both limits $f\to\pm\infty$ the free energy $\mathcal F_\tube(f)$ is asymptotic to a linear function of $f$, and we have proved the exact forms of both of these linear functions. In the $f\to-\infty$ case the asymptote can be written in terms of the growth rate of a class of objects we call full $s$-blocks; we conjecture that this value is in fact the same as the growth rate of a subclass of polygons, namely Hamiltonian polygons, which occupy all vertices within a $L\times M\times N$ rectangular prism.   {Using transfer matrix calculations related to full $s$-blocks, we establish that the conjecture is true for tube sizes
including $M=0$ and $1\leq L \leq 8$, $M=1$ and $1\leq L \leq 4$, and $(L,M)=(2,2)$.}

{Note that, if the conjecture holds, then essentially the order of the two limits $n\to\infty$ (polygon length grows to infinity) and $f\to -\infty$ (the force becomes infinitely compressive) can be interchanged.  When the conjecture is true, there is at least one consequence of this with respect to the probability of knotting.   Specifically, the properties of Hamiltonian polygons presented here in Section \ref{sec:hamiltonian},  have been used previously in \cite[Theorem 4.3]{Eng2014Thesis} to establish that: for any given proper $r$-pattern $P$ obtained from a Hamiltonian polygon in $\tube$, all but exponentially few sufficiently large Hamiltonian polygons in $\tube$ will contain $P$.    Then for $L\geq 2$, $M\geq 1$,  letting $P$ be an appropriate full tight trefoil pattern c.f. \cite[Figure 4.12]{Eng2014Thesis}, this establishes that all but exponentially few  sufficiently large Hamiltonian polygons in $\tube$ are knotted.   Combining this with the Atapour et al \cite{Atapouretal2009Stretched} results about knotting for finite forces $f$, we have that if the $f\to -\infty$ limit is dominated exponentially by Hamiltonian polygons, then for any force $f\in[-\infty, \infty)$, all but exponentially few  sufficiently large  polygons in $\tube$ will be knotted.   }

\bibliographystyle{plain}
\bibliography{tubes}

\end{document}